\documentclass[10pts,nofootinbib,superscriptaddress]{revtex4-1}
\usepackage{helvet}
\usepackage[latin9]{inputenc}
\usepackage{amsmath}
\usepackage{amssymb}
\usepackage{graphicx}
\usepackage{mathpazo}
\usepackage{wasysym}
\usepackage{epsfig}
\usepackage{amsfonts}
\usepackage{color}
\usepackage{graphics}
\usepackage{epstopdf}
\usepackage{adjustbox}
\usepackage{lipsum}\usepackage{datetime}
\usepackage{color}\usepackage{wasysym}
\usepackage{slashed}
\usepackage[unicode=true, bookmarks=false, breaklinks=false,pdfborder={0 0 1},backref=section,colorlinks=false]{hyperref}
\usepackage[justification=raggedright, margin=5mm,font=sf,small,labelfont=bf, format=plain]{caption}

\begin{document}
\title{\color{blue}Phenomenoly of the Hidden SU(2) Vector Dark Matter Model}

\author{Nabil Baouche}
\email{baouche.nabil@gmail.com}
\affiliation{Faculty of Science and Technology, University of Jijel PB 98 Ouled
Aissa, DZ-18000 Jijel, Algeria.}
\affiliation{Laboratoire de Physique des Particules et Physique Statistique, Ecole
Normale Superieure, BP 92 Vieux Kouba, DZ-16050 Algiers, Algeria.}

\author{Amine Ahriche}
\email{ahriche@sharjah.ac.ae}
\affiliation{Department of Applied Physics and Astronomy, University of Sharjah,
P.O. Box 27272 Sharjah, UAE.}
\affiliation{The Abdus Salam International Centre for Theoretical Physics, Strada
Costiera 11, I-34014, Trieste, Italy.}

\author{Gaber Faisel}
\email{gaberfaisel@sdu.edu.tr}
\affiliation{Department of Physics, Faculty of Arts and Sciences, Sueleyman Demirel
University, Isparta, Turkey 32260.}

\author{Salah Nasri}
\email{snasri@uaeu.ac.ae}
\affiliation{Department of physics, United Arab Emirates University, Al-Ain, UAE.}
\affiliation{The Abdus Salam International Centre for Theoretical Physics, Strada
Costiera 11, I-34014, Trieste, Italy.}
\begin{abstract}
We investigate the phenomenology of an extension of the Standard Model
(SM) by a non-abelian gauge group $SU(2)_{HS}$ where all
SM particles are singlets under this gauge group, and a new scalar
representation $\phi$ that is singlet under SM gauge group and doublet
under $SU(2)_{HS}$. In this model, the dark matter (DM) candidates
are the three mass degenerate dark photons $A_{i}$ $(i=1,2,3)$ of
$SU(2)_{HS}$; and the hidden sector interacts with the (SM) particles
through the Higgs portal interactions. Consequently, there will be a new
CP-even scalar $\eta$ that could be either heavier or lighter than
the SM-like Higgs. By taking into account all theoretical and experimental
constraints such as perturbativity, unitarity, vacuum stability, non-SM
Higgs decays, DM direct detection, DM relic density, we found viable
DM is possible in the range from GeV to TeV. Within the viable parameters
space, the both of the triple Higgs coupling and the di-Higgs production
at LHC14 could be enhanced or reduced depending on the scalar mixing
and the mass of the scalar particle $\eta$. 
\end{abstract}
\maketitle

\section{Introduction}

It is a fact that $27\%$ of the matter in the universe is made out
of cold dark matter (CDM). Historically, its existence was proposed
as a possible explanation for several astrophysical observations in
the cluster of galaxies\cite{Zwicky:1933gu}. 
The combined analysis of the Planck satellite 2018 results gives a
value of the relic abundance of DM density~\cite{Aghanim:2018eyx}
\begin{equation}
\Omega_{{\rm DM}}h^{2}=0.120\pm0.001\,,\label{eq:omegah}
\end{equation}
where $h$ is the reduced Hubble constant and $\Omega_{{\rm DM}}$
denotes the energy density of DM in unit of the critical density.
Obviously, the DM candidate must be a stable particle, or at least
its lifetime is much larger than the lifetime of the universe, with
no direct interaction with the electroweak and strong forces. Its
stability can be guaranteed by imposing an appropriate symmetry, which
can be discrete or continuous. In addition, it has also to be non-relativistic,
i.e. cold as the possibility of hot dark matter (DM) is ruled out
by several observations. Among these observations one can list briefly,
the pattern of fluctuations in the cosmic microwave background, the
so early formation of stars, galaxies, and clusters of galaxies and
the weak lensing signals we observe. Thus one should consider going
beyond SM physics for possible DM candidates. In the literature, many
extensions of the SM have been proposed with the DM being a scalar~\cite{Deshpande:1977rw}, fermion~\cite{Nath:1996xh} or a vector boson~\cite{Cheng:2002ej,Hambye:2008bq}. 

In this work, we consider a model, beyond SM, where DM candidates
as a multiplet of massive non-standard spin-1 gauge bosons that interact
with the SM through the Higgs portal, which was originally proposed
in~\cite{Hambye:2008bq}. This model has the privilege that the stability
of the DM particles is guaranteed by the custodial symmetry associated
to the gauge symmetry and particle content of the model. This symmetry
is favored as one, in this case, does not need to impose any kind
of discrete or global symmetry by hand. However, several important
features of the model were not investigated due to the non-observation
of the Higgs boson at the time of the study. For instances, the measurements
of some of the Higgs couplings to a reasonable precision can be used
 to constrain any heavy scalar state which mixes with the SM-like
Higgs boson which will be carried in this work. Additionally, triple
Higgs coupling and the Di-Higgs production turn to be so important
to shed light on new physics and to understand the electroweak symmetry
breaking. All these issues will be investigated in this work. Moreover,
we provide the complete explicit analytic expressions of the cross
sections of the DM annihilation and co-annihilation different channels
that contribute to the thermal relic density at freeze out needed
for estimating DM relic density. These expressions were not reported
in~\cite{Hambye:2008bq}. Finally, we will report the results related
to the branching ratios of several decay modes of the scalars in the
model which can be tested in future collider experiments.

The paper is organized as follows. First, we review the model proposed
in~\cite{Hambye:2008bq} and discuss the mass eigenstates and their
couplings that arise from the scalar potential in section~\ref{sec:Model}.
Then in section~\ref{sec:constr}, we investigate the theoretical
and experimental constraints, such as vacuum stability, unitarity,
and DM direct detection bound, that can be imposed on the model parameters.
In section~\ref{sec:DM}, we consider the lightest gauge vector of
the $SU(2)_{HS}$ to be a DM candidate and estimate its relic density
by considering all possible allowed annihilation channels as well
the coannihilation effects. Next, we carry out a detailed study of
the collider phenomenology of the model in section~\ref{sec:Pheno}.
Finally, we summarize our results and conclude in section~\ref{sec:Conc}.
Some relevant formulas and expressions of the effective potential
and the cross section contributionss are given in Appendix~\ref{sec:One-Loop}
and Appendix~\ref{sec:XSection}, respectively.

\section{Model\label{sec:Model}}

In this study, we consider the proposed model in~\cite{Hambye:2008bq}.
The model is based on enlarging the gauge symmetry of the SM to include
a non-abelian gauge symmetry, is referred to as $SU(2)_{HS}$, under
which all SM particles are singlets. The scalar sector of the model
contains a new doublet $\phi$ that is charged under the group $SU(2)_{HS}$
and is singlet under the SM gauge group. The extra gauge bosons, associated
to $SU(2)_{HS}$, are denoted by $A^{\mu}$. They couple to the SM
only through the Higgs portal ($-\lambda_{m}\phi^{\dagger}\phi H^{\dagger}H$)
and do not mix with the SM gauge bosons due to the non-abelian character
of $SU(2)_{HS}$~\cite{Hambye:2008bq}. The Lagrangian of the model
can be expressed as
\begin{equation}
{\cal L}\supset-\frac{1}{4}F^{\mu\nu}\cdot F_{\mu\nu}+(D_{\mu}\phi)^{\dagger}(D^{\mu}\phi)-\lambda_{m}\phi^{\dagger}\phi H^{\dagger}H-\mu_{\phi}^{2}\phi^{\dagger}\phi-\lambda_{\phi}(\phi^{\dagger}\phi)^{2},\label{inputlagr}
\end{equation}
where $D^{\mu}\phi=\big(\partial^{\mu}-ig_{\phi}\frac{\tau_{i}}{2}A_{i}^{\mu}\big)\phi$
with $g_{\phi}$ is the $SU(2)_{HS}$ gauge coupling; and $\tau_{i}$
are the Pauli matrices. The Higgs potential of the SM is parameterized
as: ${\cal L}^{SM}\supset-\mu^{2}H^{\dagger}H-\lambda(H^{\dagger}H)^{2}$
with $H^{T}=(\chi^{+},\frac{1}{\sqrt{2}}[h'+i\chi^{0}])$. For suitable
choice of the parameters $\mu_{\phi}^{2}$ and $\lambda_{m}$, the
gauge symmetry $SU(2)_{HS}$ is spontaneously broken when the vacuum
expectation value of $\phi$, $\upsilon_{\phi}$, is not vanishing.
After expressing the new doublet in the unitary gauge as 
\begin{equation}
\phi=\frac{\upsilon_{\phi}+\eta'}{\sqrt{2}}\exp\left\{ -i\tau_{k}\frac{\xi_{k}}{\upsilon_{\phi}}\right\} \left(\begin{array}{c}
0\\
1
\end{array}\right),
\end{equation}
one gets

\begin{eqnarray}
{\cal L} & = & {\cal L}_{SM}-\frac{1}{4}F_{\mu\nu}\cdot F^{\mu\nu}+\frac{1}{8}(g_{\phi}\upsilon_{\phi})^{2}A_{\mu}\cdot A^{\mu}+\frac{1}{8}g_{\phi}^{2}A_{\mu}\cdot A^{\mu}\eta'^{2}+\frac{1}{4}g_{\phi}^{2}\upsilon_{\phi}A_{\mu}\cdot A^{\mu}\eta'\nonumber \\
 & & +\frac{1}{2}(\partial_{\mu}\eta')^{2}-\frac{\lambda_{m}}{2}(\eta'+\upsilon_{\phi})^{2}H^{\dagger}H-\frac{\mu_{\phi}^{2}}{2}(\eta'+\upsilon_{\phi})^{2}-\frac{\lambda_{\phi}}{4}(\eta'+\upsilon_{\phi})^{4}\,,\label{lagrzerov}
\end{eqnarray}
where $A_{\mu}=UA'_{\mu}U^{-1}-\frac{i}{g_{\phi}}[\partial_{\mu}U]U^{-1}$
with $U=\exp\{-i\tau_{k}\xi_{k}/\upsilon_{\phi}\}\in SU(2)_{HS}$.
In the $A_{i}^{\mu}$ component space, the Lagrangian ${\cal L}$
in (\ref{lagrzerov}) has $SO(3)$ custodial symmetry. As a consequence,
the three $A_{i}^{\mu}$ components are degenerate in mass, $m_{A}=g_{\phi}\upsilon_{\phi}/2$,
stable and hence can serve as vector DM candidates.

In order to proceed, we need to express ${\cal L}$ in (\ref{lagrzerov})
in terms of the mass eigenstates. This can be done after minimizing
the scalar potential along both $\phi$ and $H$ directions. Setting
$H=\exp\{-i\tau_{k}\chi_{k}/\upsilon\}.(0,\,\frac{1}{\sqrt{2}}[\upsilon+h'])^{T}$,
where $\upsilon=246\,GeV$ is the usual Higgs vacuum expectation value.
By imposing the tadpole conditions $\partial V/\partial h=\partial V/\partial\eta=0$
one finds 
\begin{equation}
\mu^{2}=-\lambda\upsilon^{2}-\frac{1}{2}\lambda_{m}\upsilon_{\phi}^{2},\,\mu_{\phi}^{2}=-\lambda_{\phi}\upsilon_{\phi}^{2}-\frac{1}{2}\lambda_{m}\upsilon^{2}.\label{eq:tad}
\end{equation}
The $h'$-$\eta'$ mixing due to the presence of the term of $\lambda_{m}$
in (\ref{lagrzerov}), leads to the mass squared matrix 
\begin{equation}
M^{2}=\left(\begin{array}{cc}
2\,\lambda\,\upsilon^{2} & \lambda_{m}\,\upsilon\,\upsilon_{\phi}\\
\lambda_{m}\,\upsilon\,\upsilon_{\phi} & 2\,\lambda_{\phi}\,\upsilon_{\phi}^{2}
\end{array}\right),\label{massmatrix}
\end{equation}
which gives the eigenvalues and the mixing 
\begin{equation}
m_{1,2}^{2}=\lambda\upsilon^{2}+\lambda_{\phi}\upsilon_{\phi}^{2}\mp\sqrt{\left(\lambda\upsilon^{2}-\lambda_{\phi}\upsilon_{\phi}^{2}\right)^{2}+\lambda_{m}^{2}\upsilon^{2}\upsilon_{\phi}^{2}},\,t_{2\beta}=\frac{\lambda_{m}\,\upsilon\,\upsilon_{\phi}}{\lambda_{\phi}\,\upsilon_{\phi}^{2}-\lambda\,\upsilon^{2}},\label{eq:mHeta}
\end{equation}
where $t_{2\beta}=\tan2\beta$. By diagonalizing $M^{2}$, we get
the mass eigenstates $h$ and $\eta$ that are defined as 
\begin{equation}
\left(\begin{array}{c}
h\\
\eta
\end{array}\right)=\left(\begin{array}{cc}
c_{\beta} & -s_{\beta}\\
s_{\beta} & c_{\beta}
\end{array}\right)\,\left(\begin{array}{c}
h'\\
\eta'
\end{array}\right),
\end{equation}
with $c_{\beta}=\cos\beta,\,s_{\beta}=\sin\beta$. Since the couplings
are real, the eigenvalues of the matrix $M^{2}$ are required to be
positive definite only if 
\begin{equation}
\lambda>0,\,\lambda_{\phi}>0,\,2\lambda_{m}+\sqrt{\lambda\lambda_{\phi}}>0.\label{eq:Vstab}
\end{equation}

In our analysis, we identify the $m_{h}\sim125\,GeV$ eigenstate to
be the SM-like Higgs boson and $\eta$ the other eigenstate, therefore,
we have two cases where the SM-like Higgs eiegenstate is the (1) light
or the (2) heavy eigenstate. Then, in the first case, the CP-even
scalar masses (\ref{eq:mHeta}) should be written as 
\begin{equation}
m_{h,\eta}^{2}=\lambda\upsilon^{2}+\lambda_{\phi}\upsilon_{\phi}^{2}\mp(\lambda_{\phi}\upsilon_{\phi}^{2}-\lambda\upsilon^{2})/c_{2\beta},
\end{equation}
with $c_{2\beta}=\cos2\beta$ and the quartic couplings can be expressed
as

\begin{equation}
\lambda=\frac{m_{h}^{2}}{2\upsilon^{2}}\,c_{\beta}^{2}+\frac{m_{\eta}^{2}}{2\upsilon^{2}}\,s_{\beta}^{2},\,\lambda_{\phi}=\frac{m_{h}^{2}}{2\upsilon_{\phi}^{2}}\,s_{\beta}^{2}+\frac{m_{\eta}^{2}}{2\upsilon_{\phi}^{2}}\,c_{\beta}^{2},\,\lambda_{m}=\frac{s_{2\beta}}{2\upsilon\upsilon_{\phi}}\,\left(m_{\eta}^{2}-m_{h}^{2}\right).
\end{equation}

In the second case, the CP-even scalar masses are given by (\ref{eq:mHeta})
are 
\begin{equation}
m_{h,\eta}^{2}=\lambda\upsilon^{2}+\lambda_{\phi}\upsilon_{\phi}^{2}\pm(\lambda_{\phi}\upsilon_{\phi}^{2}-\lambda\upsilon^{2})/c_{2\beta},
\end{equation}
and the quartic couplings 
\begin{equation}
\lambda=\frac{m_{h}^{2}}{2\upsilon^{2}}\,s_{\beta}^{2}+\frac{m_{\eta}^{2}}{2\upsilon^{2}}\,c_{\beta}^{2},\,\lambda_{\phi}=\frac{m_{h}^{2}}{2\upsilon_{\phi}^{2}}\,c_{\beta}^{2}+\frac{m_{\eta}^{2}}{2\upsilon_{\phi}^{2}}\,s_{\beta}^{2},\,\lambda_{m}=\frac{s_{2\beta}}{2\upsilon\upsilon_{\phi}}\,\left(m_{h}^{2}-m_{\eta}^{2}\right).
\end{equation}

Clearly, the model can be described by the four free parameters $g_{\phi}$,
$\lambda_{\phi}$, $\mu_{\phi}$ and $\lambda_{m}$, or equivalently
by $s_{\beta},\,g_{\phi},\,m_{\eta}$ and $\upsilon_{\phi}$, in addition
to the SM parameters.

Keeping only terms relevant to our study, in the Lagrangian in the
mass eigenstates basis, we can write

\begin{eqnarray}
{\cal L} & \supset & -\frac{1}{4}F_{\mu\nu}\cdot F^{\mu\nu}+\frac{1}{2}m_{A}^{2}A_{\mu}\cdot A^{\mu}-\frac{1}{2}m_{\eta}^{2}\eta^{2}-\frac{1}{2}m_{h}^{2}h^{2}-\left(h\,c_{\beta}+\eta\,s_{\beta}\right)\,\bigg(\sum_{f}\frac{m_{f}}{\upsilon}f\overline{f}\bigg)\nonumber \\
 & + & \bigg[\frac{s_{\beta}^{2}}{2\upsilon}\eta^{2}+\frac{c_{\beta}^{2}}{2\upsilon}h^{2}+\frac{s_{2\beta}}{2\upsilon}\eta h+\eta s_{\beta}+hc_{\beta}\bigg]\bigg(\frac{2m_{W}^{2}}{\upsilon}W_{\mu}^{+}W^{-\mu}+\frac{m_{Z}^{2}}{\upsilon}Z^{\mu}Z_{\mu}\bigg)\nonumber \\
 & + & \bigg[\frac{c_{\beta}^{2}}{2\upsilon_{\phi}}\eta^{2}+\frac{s_{\beta}^{2}}{2\upsilon_{\phi}}h^{2}-\frac{s_{2\beta}}{2\upsilon_{\phi}}\eta h+\eta c_{\beta}-hs_{\beta}\bigg]\bigg(\frac{m_{A}^{2}}{\upsilon_{\phi}}A_{\mu}\cdot A^{\mu}\bigg)\nonumber \\
 & - & \frac{1}{6}\rho_{h}h^{3}-\frac{1}{6}\rho_{\eta}\eta^{3}-\frac{1}{2}\rho_{1}\eta^{2}h-\frac{1}{2}\rho_{2}h^{2}\eta.\label{lagrfin}
\end{eqnarray}
where $f$ denotes the SM fermions and the $\rho$'s parameters represents
the CP-even scalar triple coiplings that are given by 
\begin{eqnarray}
\rho_{1} & = & -6\lambda_{\phi}\upsilon_{\phi}c_{\beta}^{2}s_{\beta}+6\lambda\upsilon s_{\beta}^{2}c_{\beta}+\lambda_{m}(\upsilon c_{\beta}^{3}-\upsilon_{\phi}s_{\beta}^{3}-2\upsilon c_{\beta}s_{\beta}^{2}+2\upsilon_{\phi}c_{\beta}^{2}s_{\beta}),\nonumber \\
\rho_{2} & = & 6\lambda_{\phi}\upsilon_{\phi}s_{\beta}^{2}c_{\beta}+6\lambda\upsilon c_{\beta}^{2}s_{\beta}+\lambda_{m}(\upsilon s_{\beta}^{3}+\upsilon_{\phi}c_{\beta}^{3}-2\upsilon c_{\beta}^{2}s_{\beta}-2\upsilon_{\phi}s_{\beta}^{2}c_{\beta}),\nonumber \\
\rho_{\eta} & = & 6\lambda_{\phi}\upsilon_{\phi}c_{\beta}^{3}+6\lambda\upsilon s_{\beta}^{3}+3\lambda_{m}c_{\beta}s_{\beta}(\upsilon c_{\beta}+\upsilon_{\phi}s_{\beta}),\nonumber \\
\rho_{h} & = & -6\lambda_{\phi}\upsilon_{\phi}s_{\beta}^{3}+6\lambda\upsilon c_{\beta}^{3}+3\lambda_{m}c_{\beta}s_{\beta}(\upsilon s_{\beta}-\upsilon_{\phi}c_{\beta}).\label{eq:rho}
\end{eqnarray}

We notice from (\ref{lagrfin}), that the couplings of $h$ and $\eta$
to SM particles are weighted by $c_{\beta}$ and $s_{\beta}$, respectively.
Moreover, in the first case the scalar $\eta$ can additionally decay
into Higgs pairs if $m_{\eta}>2m_{h}$ with the partial deacy width
\begin{equation}
\varGamma\,(\eta\rightarrow hh)=\Theta(m_{\eta}-2m_{h})\frac{\rho_{2}^{2}}{32\pi m_{\eta}}\sqrt{1-4\frac{m_{h}^{2}}{m_{\eta}^{2}}},
\end{equation}
and in the second case, the Higgs can decay into a pair of $\eta$
if $m_{h}>2\,m_{\eta}$, in addition to their possible decay to $A_{i}A_{i}$.

\section{Theoretical \& Experimental Constraints\label{sec:constr}}

This model is subject to a number of theoretical and experimental constraints
such as vacuum stability, perturbativity, perturbative unitarity, electroweak
precision tests (EWPT), and the constraints from the Higgs decay. For
the EWPT, it is expected that the new physics contribution to the
oblique parameters ($\Delta S$ and $\Delta T$) is negligeable since
the scalar doublet $\eta$ is a singlet the SM gauge group. Then,
by considering the constraints from the Higgs signal strenth $\mu_{{\rm tot}}\geq0.89~{\rm at}~95\%\,{\rm CL}$~\cite{Arcadi:2019lka},
the $h-\eta$ mixing makes both $\Delta S$ and $\Delta T$ very tiny,
and all the space parameters will be allowed by the EWPT. In what
follows, we discuss the above mentioned constraints in details.
\begin{itemize}
\item \textbf{Unitarity constraints}

Possible constraints on the quartic couplings $\lambda$, $\lambda_{\phi}$
and $\lambda_{m}$, can be derived upon requiring that the amplitudes
for the scalar-scalar scattering $S_{1}S_{2}\to S_{3}S_{4}$ at high
energies respect the tree-level unitarity~\cite{Cornwall:1974km}.
Here, $S_{i}$ can be $h$ or $\eta$ for $i=1,2,3,4$. This can be
understood as, at high energies, the dominant contributions to these
amplitudes are those mediated by the quartic couplings~\cite{Arhrib:2000is}.
Denoting the eigenvalues of the scattering matrix as $\Lambda_{i}$,
the unitarity condition reads 
\begin{equation}
|\Lambda_{i}|\leq8\pi\label{unicon}
\end{equation}
In the model under concern the above bound results in the following
constraints

\begin{equation}
\lambda_{m}\leq8\pi,\,\,\,\,\lambda\leq4\pi,\,\,\,\,\lambda_{\phi}\leq4\pi,\,\,\,\,3\big(\lambda+\lambda_{\phi}\big)\pm\sqrt{9\big(\lambda-\lambda_{\phi}\big)^{2}+4\lambda_{m}^{2}}\leq8\pi.\label{eq:Unit}
\end{equation}

\item \textbf{Vacuum Stability and Perturbativity}

The quartic couplings of the scalar potential is subjected to a number of constraints to ensure that the potential is bounded from below and 
that the couplings remain perturbative as well the electroweak vacuum
to be stable all the way up to the Planck scale. For the scalar potential
to be bounded from below, the conditions $\lambda>0$, $\lambda_{\phi}>0$
and $\lambda\lambda_{\phi}>0$ must hold and for the case $\lambda_{m}<0$,
the condition $2\lambda_{m}+\sqrt{\lambda\lambda_{\phi}}>0$ must
be also satisfied. Recall that, one needs $4\lambda\lambda_{\phi}>\lambda_{m}^{2}$
so that the Higgs mixing matrix $M^{2}$ is positive definite and
thus $m_{h},m_{\eta}>0$.

\item \textbf{Constraints on the Higgs Decays}

Since the Higgs couplings are modified in our model, and since there
exist new particles with new interactions, then the Higgs total decay
width and branching ratios get modified. Here, all the vertex of the
Higgs-gauge fields and Higgs-fermions are scaled by $c_{\beta}$,
therefore the Higgs partial decay widths to the SM paritcles are scaled
as $\Gamma(h\rightarrow X_{SM}\bar{X}_{SM})=c_{\beta}^{2}\,\text{\ensuremath{\Gamma}}_{SM}(h\rightarrow X_{SM}\bar{X}_{SM})$.
In addition to the SM final states, the Higgs may decay into new gauge
bosons ( dark photons) if kinematicaly allowed, with the Higgs partial
decay width

\begin{equation}
\varGamma_{inv}\,(h\rightarrow AA)=\sum_{i=1}^{3}\varGamma\,(h\rightarrow A_{i}A_{i})=3\Theta(m_{h}-2m_{A})\frac{g_{\phi}^{2}\,m_{h}^{3}s_{\beta}^{2}}{64\pi m_{A}^{2}}\sqrt{1-4\frac{m_{A}^{2}}{m_{h}^{2}}}\left\{ 1-4\frac{m_{A}^{2}}{m_{h}^{2}}+12\frac{m_{A}^{4}}{m_{h}^{4}}\right\} ,\label{eq:Ginv}
\end{equation}
this channel is open when $m_{A}<m_{h}/2$. The factor ``3'' in
(\ref{eq:Ginv}) refers to the summation over $A_{i}A_{i}$. In Case~$2$
where the condition $m_{h}>2\,m_{\eta}$ may be fulfilled, the decay
channel $h\rightarrow\eta\eta$ is open and the partial width is given
by 
\begin{equation}
\varGamma\,(h\rightarrow\eta\eta)=\Theta(m_{h}-2m_{\eta})\frac{\rho_{1}^{2}}{32\pi m_{h}}\sqrt{1-4\frac{m_{\eta}^{2}}{m_{h}^{2}}}.\label{Gund}
\end{equation}

Therefore, the Higgs total decay width can be written as 
\begin{equation}
\Gamma_{h}=\text{\ensuremath{\Gamma_{BSM}}}+c_{\beta}^{2}\,\ensuremath{\Gamma}_{h}^{SM},\label{eq:Cases}
\end{equation}
where $\ensuremath{\Gamma}_{h}^{SM}=4.2~\mathrm{MeV}$ is the Higgs
total decay width in the SM; $\Gamma_{BSM}=\text{\ensuremath{\Gamma_{inv}}}$
for Case$\,1$; and $\text{\ensuremath{\Gamma_{BSM}}}=\text{\ensuremath{\Gamma_{inv}}+\ensuremath{\Gamma_{und}}}$
for Case$\,2$. Here, the undetermined Higgs decay width $\Gamma_{und}=\Gamma(h\rightarrow\eta\eta)$,
which is different than the invisible one at collider; since the light
scalar can be seen at detector via the decay to light fermions $\eta\rightarrow f\bar{f}$.
These decays do not match the known SM ones, hence, the signal $h\rightarrow\eta\eta\rightarrow f_{1}\bar{f_{1}}f_{2}\bar{f_{2}}$
is named untermined. Then, the invisible and undetermined branching
ratio must respect the constarints~\cite{Aad:2019mbh,Aaboud:2019rtt}\footnote{Recent measurments by ATLAS~\cite{ATLAS:2020kdi} and CMDS~\cite{Sirunyan:2018owy} of the invisible decay width of the SM
Higgs boson give $\mathcal{B}_{inv}<0.11$ and $\mathcal{B}_{inv}<0.19$, respectively. In our numerical scan, we will consider the recent ATLAS bound.}
\begin{equation}
\mathcal{B}_{inv}<0.26,\,\mathcal{B}_{und}<0.22,\,\mathcal{B}_{inv}+\mathcal{B}_{und}\leq0.47.\label{eq:Bh}
\end{equation}
In addition, the Higgs total decay width should lie in the range~\cite{Khachatryan:2015mma,Aaboud:2018puo}
\begin{equation}
1.0~\mathrm{MeV}<\Gamma_{h}<6.0~\mathrm{MeV}.\label{eq:Gh}
\end{equation}

\item \textbf{DM Direct Detection Constraints}

The DM candidate can interact with nucleons via $t$-channel diagrams
that are mediated by h and $\eta$ as shown in Fig.~\ref{DD}. 

\begin{figure}[!h]
\includegraphics[width=0.4\textwidth]{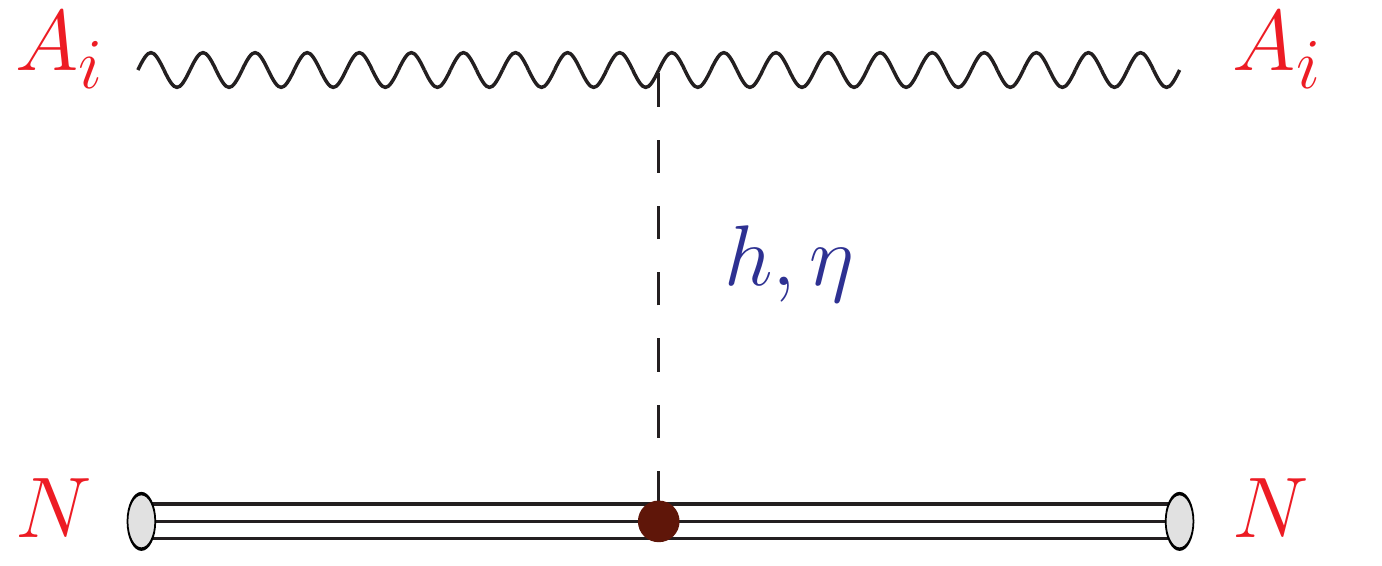} \caption{Feynman diagrams that lead to DM direct detection at underground detectors.}
\label{DD} 
\end{figure}

Measuring nuclear recoil energy resulting from the elastic scattering
of DM particle off nucleus in detector can serve as a direct search
of DM particles. The results of such searches can be used to impose
constraints on the relevant parameters of the model. At tree-level,
the spin independent elastic scattering of the vector DM off nucleus,
mediated by $h$ or $\eta$ exchange, given in~\cite{Hambye:2008bq},
can be simplifed as 
\begin{align}
\sigma_{SI}\left(NA\rightarrow NA\right) & =\frac{1}{64\pi}f^{2}g_{\phi}^{4}\,s_{2\beta}^{2}m_{N}^{2}\,\frac{\upsilon_{\phi}^{2}}{\upsilon^{2}}\frac{(m_{h}^{2}-m_{\eta}^{2})^{2}}{m_{h}^{4}m_{\eta}^{4}}\left(\frac{m_{N}}{m_{N}+m_{A}}\right)^{2}\nonumber \\
 & =6.45765608\times10^{-42}cm^{2}\,s_{2\beta}^{2}\left(\frac{g_{\phi}}{0.5}\right)^{4}\left(\frac{\upsilon_{\phi}}{100\,\mathrm{GeV}}\right)^{2}\left(\frac{m_{h}}{m_{\eta}}\right)^{4}\left(1-\frac{m_{\eta}^{2}}{m_{h}^{2}}\right)^{2}\left(1+\frac{m_{A}}{m_{N}}\right)^{-2}.\label{eq:DD}
\end{align}
Here, $m_{N}$ denotes the nucleon mass and $f=0.3$~\cite{Hambye:2008bq},
parametrizes the Higgs nucleon coupling. This has to be compared with
the present experimental upper bound on this cross section~\cite{Aprile:2015uzo}.

\item \textbf{Renormalization Group Equation}

The constraints from vacuum stability and couplings perturbativity
can be determined by the renormalisation group evolution of $\lambda,\lambda_{m},\lambda_{\phi}$.
At one-loop $\beta$ functions and upon neglecting all the Yukawa
couplings except for $y_{t}$, the relevant equations are given as~\cite{Falkowski:2015iwa,Arcadi:2019lka}:
\begin{eqnarray}
16\pi^{2}d\lambda/dt & = & 24\lambda^{2}+2\lambda_{m}^{2}-6y_{t}^{4}+\lambda\,(12y_{t}^{2}-\frac{9}{5}g_{1}^{2}-9g_{2}^{2})+\frac{27}{200}g_{1}^{4}+\frac{9}{20}\,g_{1}^{2}g_{2}^{2}+\frac{9}{8}g_{2}^{4}\;,\nonumber \\
16\pi^{2}d\lambda_{\phi}/dt & = & 24\lambda_{\phi}^{2}+2\lambda_{m}^{2}-9\lambda_{\phi}g_{\phi}^{2}+\frac{9}{4}g_{\phi}^{4}\;,\nonumber \\
16\pi^{2}d\lambda_{m}/dt & = & \lambda_{m}\,(6y_{t}^{2}+12\lambda+12\lambda_{\phi}+4\lambda_{m}-\frac{9}{10}g_{1}^{2}-\frac{9}{2}g_{2}^{2}-\frac{9}{2}g_{\phi}^{2})\;,\nonumber \\
16\pi^{2}dg_{i}/dt & = & b_{i}\,g_{i}^{3}\quad\textrm{with}\quad(b_{1},b_{2},b_{3},b_{\phi})=(41/6,-19/6,-7,-43/6)\;,\nonumber \\
16\pi^{2}d\lambda_{t}/dt & = & y_{t}\,(\frac{9}{2}y_{t}^{2}-\frac{17}{20}g_{1}^{2}-\frac{9}{4}g_{2}^{2}-8g_{3}^{2})\;,\label{RGE}
\end{eqnarray}
where $g_{i}=(g_{1},g_{2},g_{3})$ represent the SM gauge couplings
and $g_{\phi}$ is the gauge coupling of the new $SU(2)_{H}$. In what
follows, we will use (\ref{RGE}) to check whether the conditions
of the vacuum stability, perutrbativity and unitarity are filfilled
at higher scales $\Lambda=100\,TeV,\,10^{4}\,TeV$ and $\Lambda=m_{Planck}$.
\end{itemize}

\section{Dark Matter Relic Density\label{sec:DM}}

In order to estimate the relic density, one has to estimate the freeze-out
temperature and the annihilation cross section. The thermal relic
density at freeze out is given in the terms of the thermally averaged
annihilation cross section by~\cite{Srednicki:1988ce}: 

\begin{equation}
\varOmega_{DM}h^{2}\backsimeq\frac{1.04\times10^{9}}{M_{pl}}\frac{x_{F}}{\sqrt{g_{*}(x_{F})}}\frac{3}{<\sigma(AA)\upsilon_{r}>},\label{relic}
\end{equation}
where $\upsilon_{r}$ is the relative velocity, $M_{pl}=1.22\times10^{19}\,GeV$
is the Plank mass, $g_{*}$ counts the total number of relativistic
degrees of freedom, and $x_{F}=m_{A}/T_{f}$ is the inverse freeze-out
temperature. The factor ``$3$'' in (\ref{relic}) comes from the
fact that the total relic density is the summation of the contributions
of the three vector bosons $A_{i}$, that have same masses, same interactions,
and hence give the same contribution to the relic density. The total
thermally averaged annihilation cross section 
\begin{equation}
<\sigma(AA)\upsilon_{r}>=\frac{1}{8Tm_{A}^{4}K_{2}^{2}(m_{A}/T)}\sum_{X}\int_{4m_{A}^{2}}^{\infty}ds~\sigma_{AA\rightarrow X}(s)\sqrt{s}\left(s-4m_{A}^{2}\right)K_{1}(\sqrt{s}/T),
\end{equation}
where $K_{1,2}$ are the modified Bessel functions and $\sigma_{AA\rightarrow X}(s)$
is the annihilation cross section due to the channel $AA\rightarrow X$,
at the CM energy $\sqrt{s}$.

The freeze-out parameter $x_{F}$ can be obtained iteratively from
the equation:

\begin{equation}
x_{F}=\log\left(\frac{5}{4}\sqrt{\frac{45}{8}}\frac{3}{2\pi^{3}}\frac{M_{pl}m_{A}<\sigma(AA)\upsilon_{r}>}{\sqrt{g_{*}(x_{F})x_{F}}}\right),\label{xf}
\end{equation}
where the factor ``$3$'' in (\ref{xf}) refers to the DM degrees
of freedom.

In case where the DM candidate is close in mass to other species,
the coannihilation effect becomes important, and in order to take
it into account, the thermal average cross section, $<\sigma(AA)\upsilon_{r}>$,
in (\ref{relic}) and (\ref{xf}) should be replaced the effective
one $\left\langle \sigma_{eff}(AA)\upsilon_{r}\right\rangle $. The
effective thermally averaged annihilation cross section is given by~\cite{Srednicki:1988ce}: 

\begin{equation}
\begin{array}{c}
\sigma_{eff}(x)=\sum_{ij}^{N}\frac{g_{i}g_{j}}{g_{eff}^{2}}\sigma_{ij}(1+\Delta_{i})^{3/2}(1+\Delta_{j})^{3/2}\exp[-x(\Delta_{i}+\Delta_{j})],\\
g_{eff}(x)=\sum_{i=1}^{N}g_{i}(1+\Delta_{i})\exp(-x\Delta_{i}),
\end{array}
\end{equation}
where $\Delta_{i}=\frac{m_{i}-m_{DM}}{m_{DM}}$ and $g_{i}$ is the
multiplicity if the species $"i"$. In our model, we consider the co-annihilation
effect coming from the mass degeneracy between the vector bosons $A_{i}$,
and therefore $g_{i}=3$, $g_{eff}=3$, $\sigma(A_{i}A_{j\ne i})=\sigma(A_{1}A_{2})$
and $\sigma(A_{i}A_{i})=\sigma(A_{1}A_{1})$. Thus, the effective
thermally averaged cross section reads:

\begin{equation}
\sigma_{eff}(AA)=\frac{1}{3}[\sigma(A_{1}A_{1}\rightarrow all)+2\sigma(A_{1}A_{2}\rightarrow all)].\label{eff}
\end{equation}

In our analysis, we estimate the relic density by considering (\ref{relic}),
(\ref{xf}) and (\ref{eff}), and confront it to the recent precise
measurements from the PLANCK satellite shown in (\ref{eq:omegah}).
Here, we consider $3\sigma$ range, i.e., $0.117\le\Omega_{DM}h^{2}\le0.123$.
In the rest of this section, we estimate the different contributions
to the cross sections $\sigma(A_{1}A_{1})$ and $\sigma(A_{1}A_{2})$.

\begin{figure}[!h]
\includegraphics[width=0.95\textwidth]{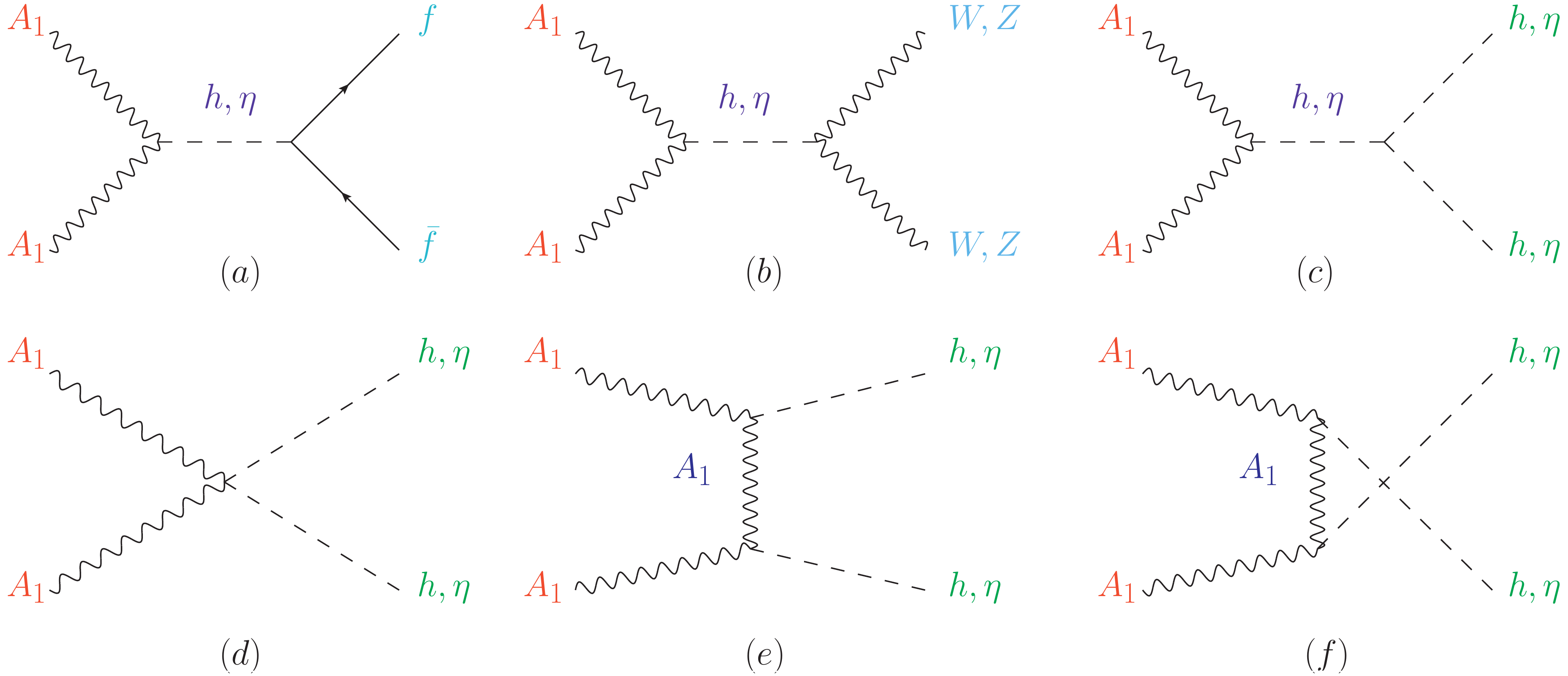} \caption{Different DM self-annihilation channels.}
\label{s1} 
\end{figure}

For the cross section $\sigma(A_{1}A_{1})$, we have many channels:
$f\bar{f}$, $WW$, $ZZ$, $hh$, $\eta\eta$ and $h\eta$ as shown
in Fig.~\ref{s1}. 

\begin{figure}[!h]
\includegraphics[width=0.66\textwidth]{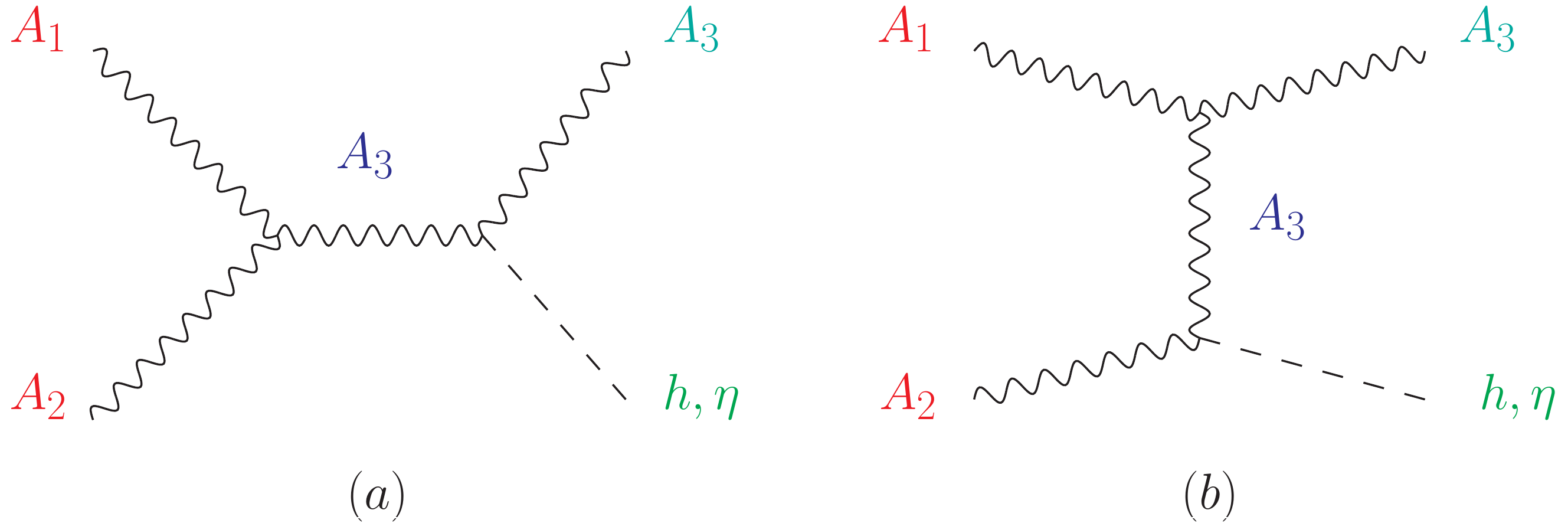} \caption{The DM co-annihilation channel.}
\label{s2} 
\end{figure}

While, for the cross section $\sigma(A_{1}A_{2})$, we have only one
channel as shown in Fig.~\ref{s2}. 

From the diagram in Fig.~\ref{s1}-a, the annihilation cross section
into fermion pairs ($f\bar{f}$) at CM enegy $\sqrt{s}$ is estimated
to be 
\begin{align}
\sigma(A_{1}A_{1}\rightarrow bf\bar{f})\upsilon_{r} & =N_{c}\frac{s_{2\beta}^{2}}{144\pi}\frac{m_{A}^{4}m_{f}^{2}}{\upsilon_{\phi}^{2}\upsilon^{2}}\,\mathcal{BW}\,\left(1-\frac{4m_{f}^{2}}{s}\right)^{\frac{3}{2}}\left(12-4\frac{s}{m_{A}^{2}}+\frac{s^{2}}{m_{A}^{4}}\right),\label{eq:ff}\\
\mathcal{BW} & =\left|\frac{i}{s-m_{h}^{2}+im_{h}\Gamma_{h}}-\frac{i}{s-m_{\eta}^{2}+im_{\eta}\Gamma_{\eta}}\right|^{2},
\end{align}
with $s_{2\beta}=\sin(2\beta)$, $N_{c}$ is the color factor (1 for
leptons and 3 for quarks), $\Gamma_{h}$ ($\Gamma_{\eta}$) is the
total decay width of the Higgs (scalar $\eta$).

The annihilation cross section into gauge bosons $V=W,Z$ as shown
in Fig.~\ref{s1}-b is found at CM enegry $\sqrt{s}$ to be 
\begin{align}
\sigma(A_{1}A_{1} & \rightarrow VV)\upsilon_{r}=\delta_{V}\frac{s_{2\beta}^{2}m_{A}^{4}m_{V}^{4}}{288\pi s\upsilon^{2}\upsilon_{\phi}^{2}}\,\mathcal{BW}\,\left(1-\frac{4m_{V}^{2}}{s}\right)^{1/2}\left(12-4\frac{s}{m_{A}^{2}}+\frac{s^{2}}{m_{A}^{4}}\right)\nonumber \\
 & \times\left(12-4\frac{s}{m_{V}^{2}}+\frac{s^{2}}{m_{V}^{4}}\right),\label{eq:VV}
\end{align}
where $V$ stands for $W^{\pm}$ or $Z$ bosons; and $\delta_{W}=2$
and $\delta_{Z}=1$.

In order to estimate the cross section for the processes $A_{1}A_{1}\rightarrow X\equiv hh,\,\eta\eta$;
we estimated the amplitude by considering the diagrams in Fig.~\ref{s1}-c,
Fig.~\ref{s1}-d, Fig.~\ref{s1}-e and Fig.~\ref{s1}-f; and write
its averaged squared in powers of $t-m_{A}^{2}$ and $u-m_{A}^{2}$,
as 
\begin{equation}
\left|\mathcal{\bar{M}}\right|^{2}=\frac{1}{2}\left\{ \frac{Q_{1}m_{A}^{4}}{(t-m_{A}^{2})^{2}}+\frac{Q_{2}m_{A}^{2}}{t-m_{A}^{2}}+Q_{3}+\frac{Q_{4}}{m_{A}^{2}}(t-m_{A}^{2})+\frac{Q_{5}}{m_{A}^{4}}(t-m_{A}^{2})^{2}+\frac{Q_{6}}{m_{A}^{6}}(t-m_{A}^{2})^{3}\right\} +\left\{ t\rightarrow u\right\} .\label{M2}
\end{equation}

Then, the integration over the angles of the $t$-terms and $u$-terms
in (\ref{M2}) leads to identical contributions, where the cross section
can be presented as 
\begin{equation}
\sigma\upsilon_{r}=\frac{Q_{0}}{s}\left\{ \frac{2Q_{1}}{A^{2}-B^{2}}+\frac{Q_{2}}{B}\ln\left(\frac{A+B}{A-B}\right)+2Q_{3}+2Q_{4}A+\frac{2}{3}Q_{5}(3A^{2}+B^{2})+2Q_{6}A(A^{2}+B^{2})\right\} ,\label{XS}
\end{equation}
where the dimensionless parameters $Q_{i}$, $A$ and $B$ are given
in Appendix~\ref{sec:XSection}. For the processes $A_{1}A_{1}\rightarrow X\equiv h\eta$
(i.e., diagrams in Fig.~\ref{s1}-c, Fig.~\ref{s1}-d and Fig.~\ref{s1}-e);
and $A_{1}A_{2}\rightarrow X\equiv A_{3}h,\,A_{3}\eta$ (i.e., diagrams
in Fig.~\ref{s2}); the averaged squared amplitude can be written
only in powers of $t-m_{A}^{2}$, i.e., 
\begin{equation}
\left|\mathcal{\bar{M}}\right|^{2}=\frac{Q_{1}m_{A}^{4}}{(t-m_{A}^{2})^{2}}+\frac{Q_{2}m_{A}^{2}}{t-m_{A}^{2}}+Q_{3}+\frac{Q_{4}}{m_{A}^{2}}(t-m_{A}^{2})+\frac{Q_{5}}{m_{A}^{4}}(t-m_{A}^{2})^{2}+\frac{Q_{6}}{m_{A}^{6}}(t-m_{A}^{2})^{3},\label{MM2}
\end{equation}
and after integration we get the cross section that has the form
(\ref{XS}). The dimensionless parameters $Q_{i}$, $A$ and $B$
for the processes $A_{1}A_{1}\rightarrow X\equiv h\eta$; and $A_{1}A_{2}\rightarrow X\equiv A_{3}h,\,A_{3}\eta$
are also given in Appendix~\ref{sec:XSection}.

\section{Colliders Phenomenology\label{sec:Pheno}}

\subsection{Collider Constraints \& the Parameters Space}

In model under consideration, we have only four free parameters. It
is convenient to choose them as: the mass of the DM, $m_{A}$, the
mass of the new scalar, $m_{\eta}$, the gauge coupling $g_{\phi}$
and finally sine of the mixing angle $\,s_{\beta}$. In our analysis,
we perform a numerical scan over the parameters space in the ranges
given as 
\begin{equation}
\left|s_{\beta}\right|\leq1,\,g_{\phi}\leq\sqrt{4\pi},m_{A},m_{\eta}\leq3\,TeV.\label{eq:SP}
\end{equation}
These ranges of the parameters space are subjected to the constraints
of vacuum stabilty (\ref{eq:Vstab}), perturbativity (at the weak
scale), unitarity (\ref{eq:Unit}), Higgs decay (\ref{eq:Bh}) and
(\ref{eq:Gh}), DM relic density (\ref{relic}), DM direct detection
(\ref{eq:DD}); in addition to the collider constraints on the Higgs
signal strength (\ref{eq:muXX}).

At the LHC, ATLAS and CMS experiments have observed the Higgs boson
in several decay channels, mainly $h\to ZZ,\,WW,\,\gamma\gamma,\,\tau\tau,\,b\bar{b}$~\cite{Khachatryan:2016vau}.
The observation allowed both collaborations to measure some of the
Higgs couplings to a reasonable precision~\cite{Khachatryan:2016vau}.
In return, these measurements can be used to constrain any heavy
scalar state which mixes with the SM--like Higgs boson. The desired
constraints can be deduced using the data of the signal strength modifier
$\mu_{XX}$ for a given search channel, $h\to XX$. The signal strength
modifier is a measured experimental quantity for the combined production
and decay and is defined as the ratio of the measured Higgs boson
rate to its SM prediction~\cite{Khachatryan:2016vau,Arcadi:2019lka}

\begin{equation}
\mu_{XX}=\sigma(pp\to h\to XX)/\sigma(pp\to h\to XX)|_{{\rm SM}}\,,
\end{equation}
in the narrow width approximation, $\mu_{XX}$ takes the simple form
as 
\begin{equation}
\mu_{XX}=\frac{\sigma(pp\to h)\times\mathcal{B}(h\to XX)}{\sigma(pp\to h)|_{{\rm SM}}\times\mathcal{B}(h\to XX)|_{{\rm SM}}}.\label{eq:muxx}
\end{equation}

As mentioned above, due to the Higgs mixing, the couplings of the
observed $h_{125}$ boson to SM fermions and gauge bosons are modified
with respect to the SM, by $c_{\beta}$. So, by considering the definitions
of the Higgs decay width in (\ref{eq:Cases}), one can simplify signal
strength modifier as 
\begin{equation}
\mu_{XX}=c_{\beta}^{2}(1-\mathcal{B}_{BSM}),\label{eq:muXX}
\end{equation}
where $\mathcal{B}_{BSM}=\mathcal{B}_{inv}$ for the case $m_{\eta}>m_{h}/2$,
and $\mathcal{B}_{BSM}=\mathcal{B}_{inv}+\mathcal{B}_{und}$ for $m_{\eta}<m_{h}/2$.
Thus, the measurement of $\mu_{XX}$ of the $h_{125}$ Higgs boson
can be used to derive a strong constraint on the mixing angle $\beta$
and the exotic Higge decay fraction. The constraints in (\ref{eq:muXX})
are complementary to those on the exotic Higgs decays shown in (\ref{eq:Bh})
and (\ref{eq:Gh}).

The ATLAS and CMS collaborations have presented the results of $\mu_{XX}$
for the various final states at the Run-I of the LHC. The LHC Run-I
corresponds to $\sqrt{s}=7$ and $\sqrt{s}=8\,TeV$ with the integrated
luminosity $5\,fb^{-1}$ and $20\,fb^{-1}$ , respectively. The combined
results have been reported in~\cite{Khachatryan:2016vau}. Based
on the reported result of the combined total signal strengths of the
$h_{125}$ Higgs, with all production and decay channels combined,
one obtains 
\begin{equation}
\mu_{{\rm tot}}\geq0.89,\label{eq:mux}
\end{equation}
at $95\%\,{\rm CL}$~\cite{Arcadi:2019lka}. This can be translated
into a bound on the Higgs mixing angle $\beta$ that reads $s_{\beta}^{2}\leq0.11$
in the absence of exotic Higgs decays, i.e., $\mathcal{B}_{BSM}=0$.
Within the LHC Run-II at $\sqrt{s}=13$ TeV, ATLAS and CMS collaborations
reported more accurate results in some decay channels. For instances,
observations at $5\sigma$ have been achieved in the $b\bar{b}$ mode~\cite{Sirunyan:2018kst,Aaboud:2018zhk}.
The combined results of $\mu_{XX}$ at $\sqrt{s}=13\,TeV$ have been
reported by CMS collaboration in~\cite{Sirunyan:2018koj}, and by
ATLAS collaboration in~\cite{Aad:2019mbh}. Unfortunately, a global
combination of their obtained results for LHC Run-II was not yet performed.
Finally, with more improvement of the experimental sensitivity on
the Higgs signal strengths in the future, a more stringent bound $s_{\beta}<0.18$
is expected to be reached at the high luminosity LHC in case of $\mathcal{B}_{BSM}=0$~\cite{Cepeda:2019klc,Arcadi:2020jqf}.
For the scenario $m_{\eta}>m_{h}$, the results of the direct searches
at the LHC, which have been performed by the ATLAS and CMS collaborations,
of heavy Higgs decays into $ZZ,WW,\gamma\gamma,hh,t\bar{t}$ can set
additional complementary constraints. In particular, the constraints
will be imposed on $m_{\eta}$ and $|s_{\beta}|$. For detailed discussion
about these constraints we refer to~\cite{Arcadi:2019lka} where
detailed study and analysis have been performed.

\begin{figure}[h!]
\begin{centering}
\includegraphics[width=0.48\textwidth]{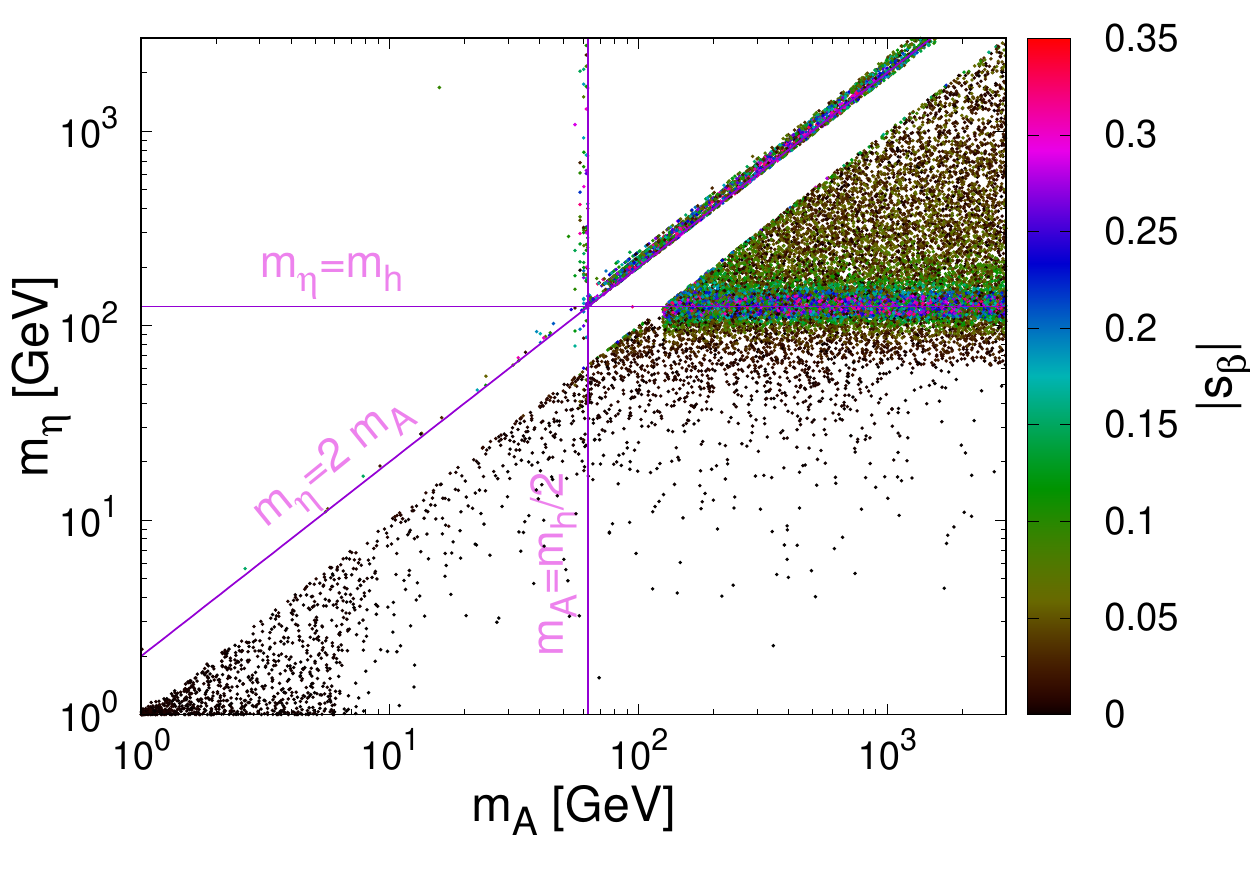}~\includegraphics[width=0.48\textwidth]{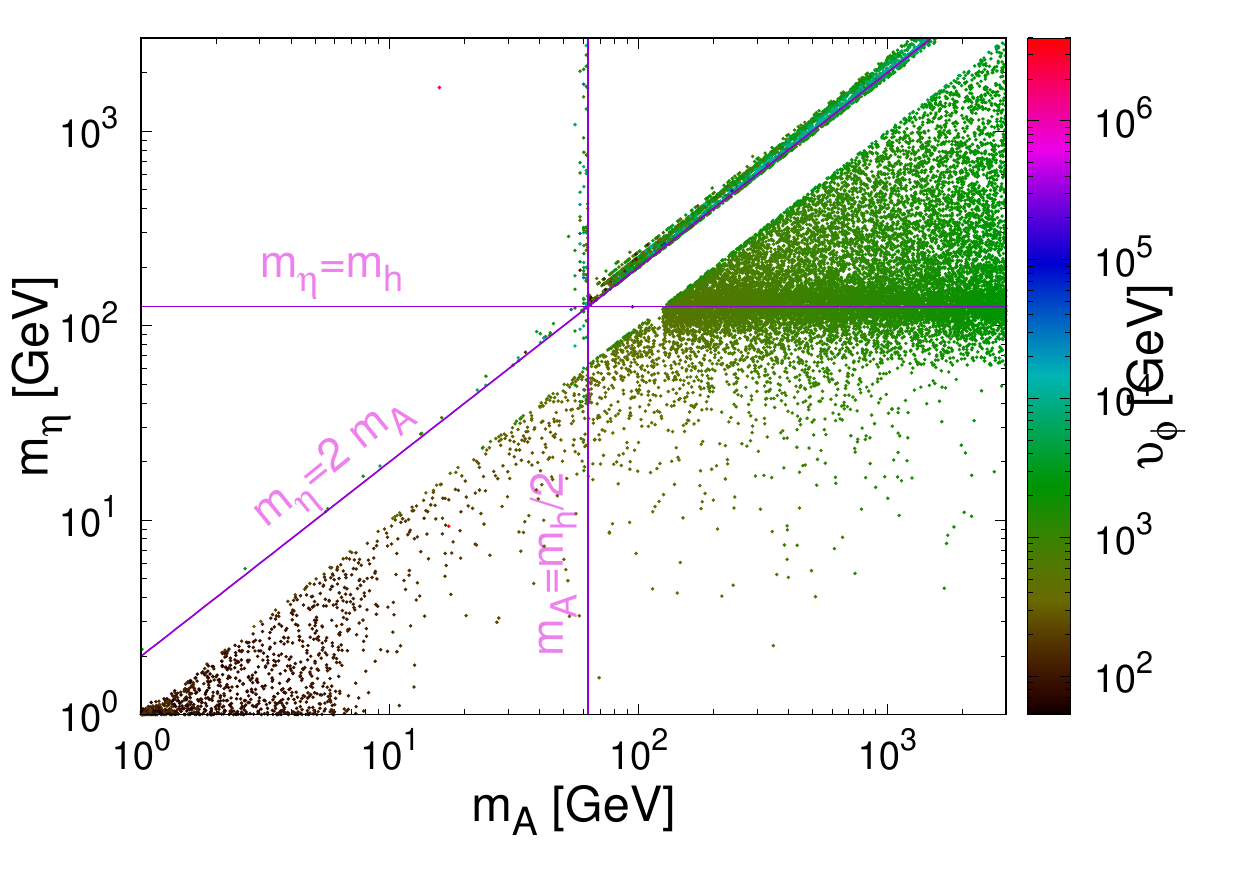}\\
 \includegraphics[width=0.48\textwidth]{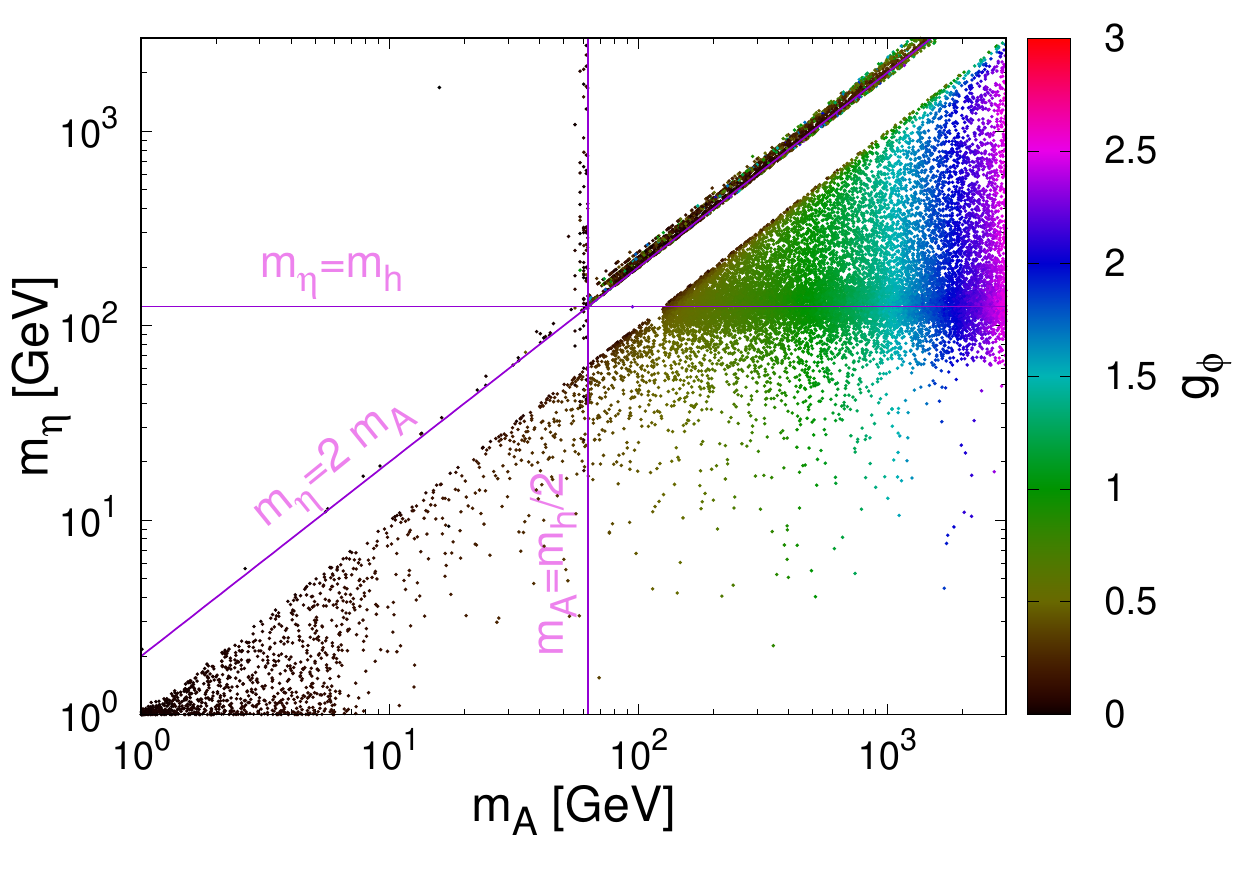}~\includegraphics[width=0.48\textwidth]{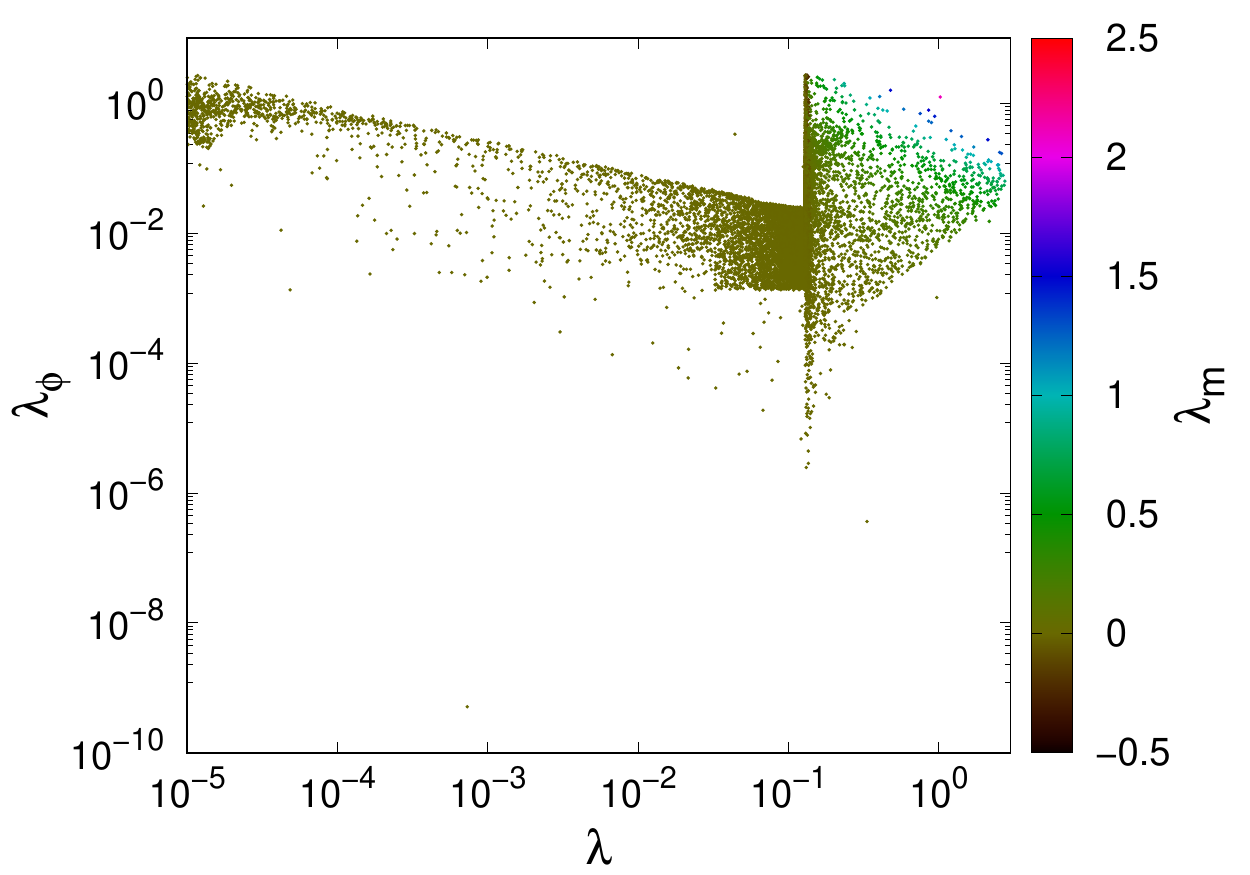} 
\par\end{centering}
\caption{The new scalar mass $m_{\eta}$ versus the DM mass $m_{A}$, where
the palette shows the scalar mixing (top-left), the new scalar vev
(top-right) and the new gauge coupling (Bottom-left). Bottom-right:
The quartic coupling $\lambda_{\phi}$ versus $\lambda$, where the
coupling $\lambda_{m}$ is shown in the palette. All the above mentioned
constraints in section~\ref{sec:constr} are taken into account during
the scan over the parameters space.}
\label{fig:SP} 
\end{figure}

\begin{figure}[h!]
\begin{centering}
\includegraphics[width=0.48\textwidth]{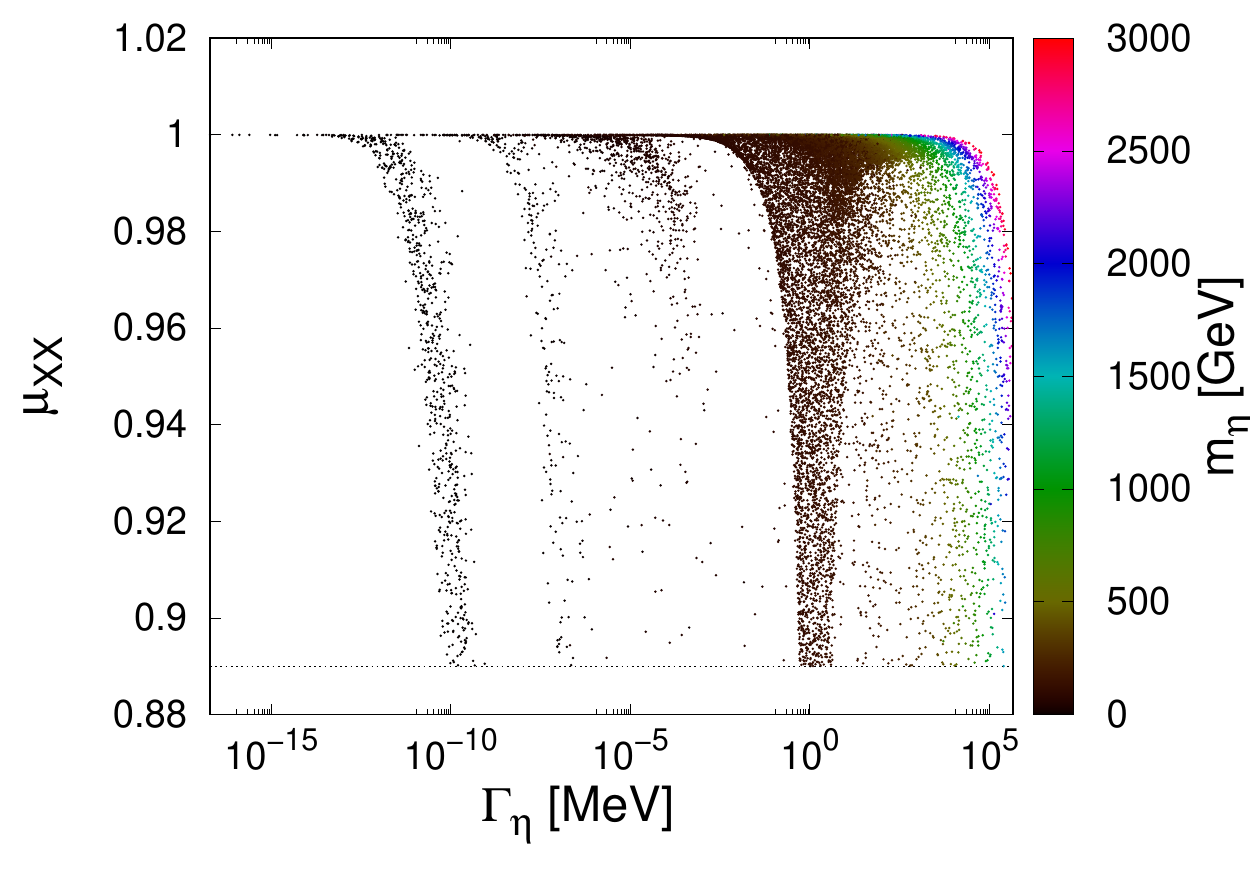}~\includegraphics[width=0.48\textwidth]{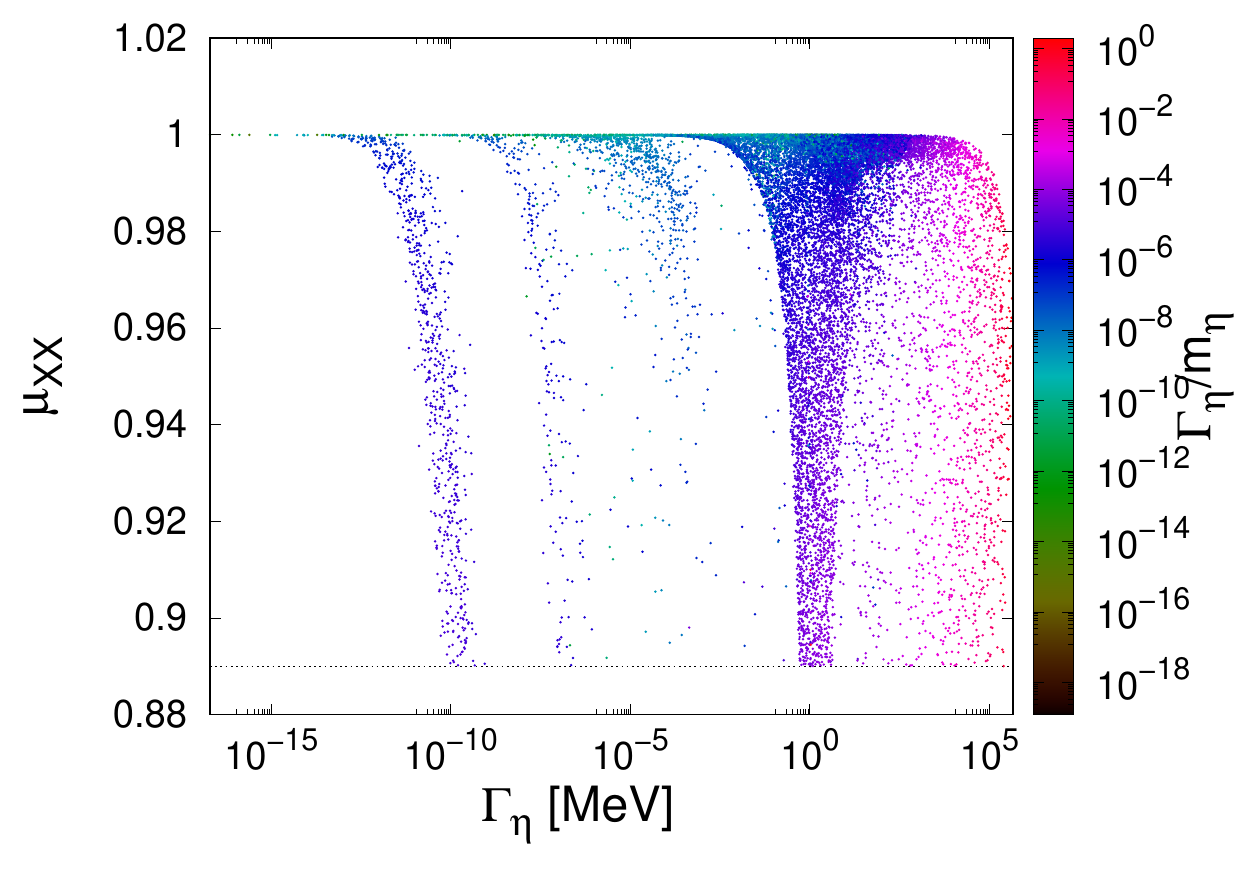} 
\par\end{centering}
\caption{The signal strength (\ref{eq:muxx}) versus the total decay width
of the new scalar ($\Gamma_{\eta}$ ). The palette shows the scalar
mass ($m_{\eta}$ in GeV) mixing in left and the ratio $\Gamma_{\eta}/m_{\eta}$
in right side.}
\label{fig:muxx} 
\end{figure}

\begin{figure}[h!]
\begin{centering}
\includegraphics[width=0.48\textwidth]{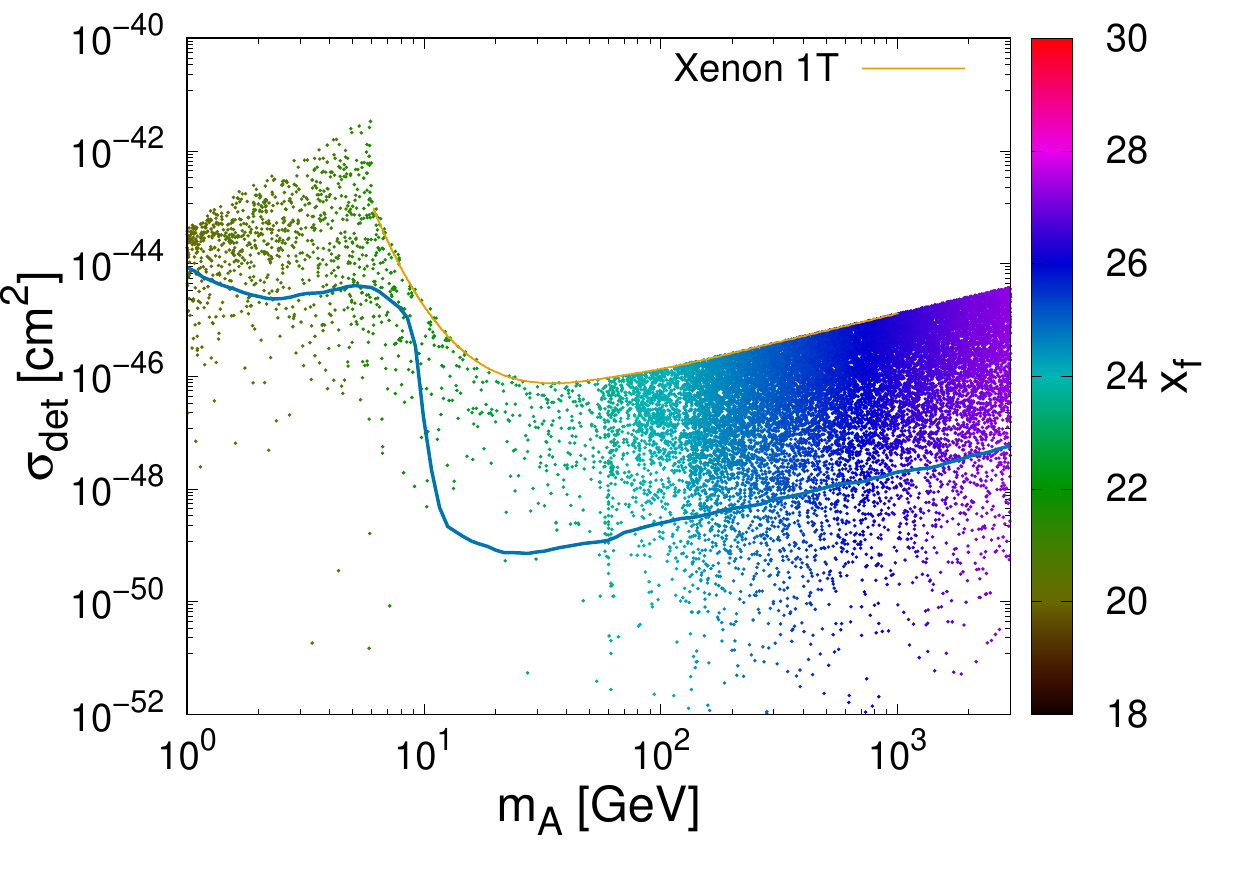}~\includegraphics[width=0.48\textwidth]{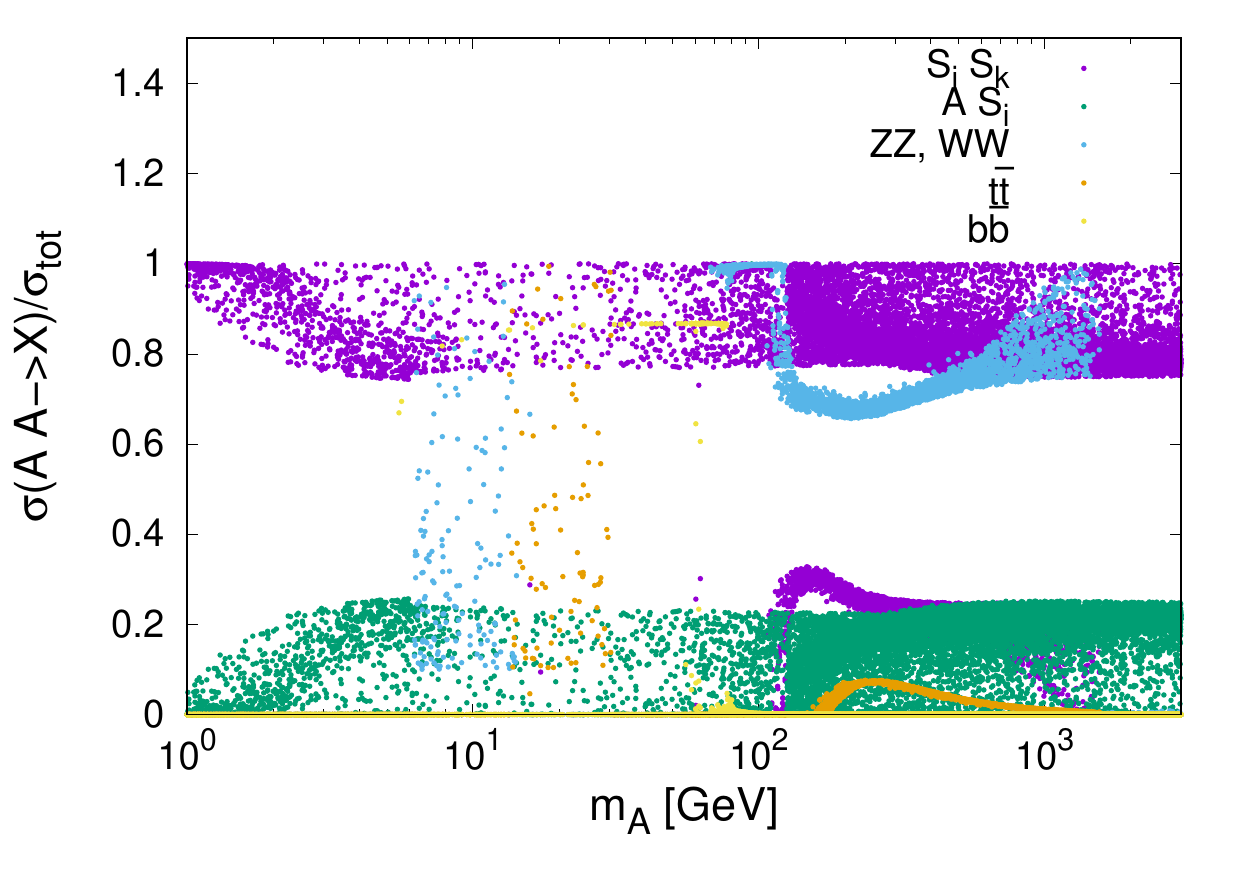} 
\par\end{centering}
\caption{Left: the DM direct detection cros section versus the DM mass, where
the palette shows the freeze-out parameter $x_{f}=m_{A}/T_{f}$. The
orange line represents the recent Xenon-1T bound~\cite{Aprile:2015uzo},
and the blue one represents the neutrino floor~\cite{Billard:2013qya}.
Right: the relative contributions to the thermal averaged cross section
of different annihilation channels.}
\label{fig:DM} 
\end{figure}

In what follows, we will show the results of our numerical scan by
considering all the above aforementioned constraints. For instance,
in Fig.~\ref{fig:SP}, we present the viable parameters space among
the intervals (\ref{eq:SP}). In Fig.~\ref{fig:muxx}, the signal
strength (\ref{eq:muxx}) is presented versus the total decay width
of the new scalar ($\Gamma_{\eta}$ ), and the DM observables are
presented in Fig.~\ref{fig:DM}.

In the top-left, top-right and bottom-left of Fig.~\ref{fig:SP},
we show the viable $m_{\eta}-m_{A}$ space parameters, respecting
the aforementioned constraints in the ranges listed in (\ref{eq:SP}),
where the palettes show the scalar mixing (top-left), the new scalar
vev $\upsilon_{\phi}$ (top-right); and the new gauge coupling $g_{\phi}$
(bottom-left). One notices the existence of three distinct sub-regions
in the $m_{\eta}-m_{A}$ plan, corresponding to (1) 
$m_{\eta}<m_{A}$, and it is the largest sub-region, (2) $m_{\eta}\gtrsim2m_{A}$, and 
(3) $m_{A}\lesssim m_{h}/2$. Moreover, we remark that the constraints imposed on the parameters
space can be easily evaded for values of $m_{\eta}$ and $m_{A}$
close to or larger than the electorweak scale, mostly in the top-right
region in each plot, for the preferred ranges of the other parameters
given as $\mid s_{\beta}\mid\lesssim0.1$, $\upsilon_{\phi}\gtrsim10^{3}$
and large values of $g_{\phi}\gtrsim0.1$. One has to mention that
most of the non-viable parameters space (empty regions in Fig.~\ref{fig:SP}-top-right
for example) are excluded mainly by DM direct detection and relic
density requirements.

Recall that in setup, the new physics contribution to the oblique
parameters $\Delta S$ and $\Delta T$ has negligible effect. As a consequence,
having small values of $s_{\beta}$ (within the condition (\ref{eq:mux}))
allows to have values $m_{\eta}$ close to or larger than the electorweak
scale without violating the constraints imposed on $\Delta S$ and
$\Delta T$. On the other hand, DM direct detection constraints given
in (\ref{eq:DD}), can be respected for large values of DM masses.
Concerning the DM mass, that is given $m_{A}=g_{\phi}\upsilon_{\phi}/2$,
having a small values of $g_{\phi}$ can be compensated by large values
$\upsilon_{\phi}$ to get the desired values of $m_{A}$ that explain
the parameters space in Fig.~\ref{fig:SP}. One remarks that the
scalar mixing is almost suppressed except in the regions around the
degeneracies that are defined by the stright lines in Fig.~\ref{fig:SP},
i.e., $m_{\eta}\gtrsim2m_{A}$, $m_{A}\lesssim m_{h}/2$ and $m_{\eta}\thicksim m_{h}$.

In Fig.~\ref{fig:SP} bottom-right, we show the quartic coupling
$\lambda_{\phi}$ versus $\lambda$ for the values of the coupling
$\lambda_{m}$ shown in the palette. One notes that the constraints
can be escaped when the mixing coupling $\lambda_{m}$ is small, preferably
less than $0.5$, and the other quartic couplings $\lambda_{\phi}$
and $\lambda$ have values approximately greater than $10^{-3}$.
This result is a consequence of applying the unitarity constraints
given by the last inequality in (\ref{eq:Unit}). Clearly, large
values of both $\lambda_{\phi}$ and $\lambda$ should be accompanied
by small values of $\lambda_{m}$ to satisfy the unitarity constraints.

In Fig.~\ref{fig:muxx}, we show the signal strength (\ref{eq:muxx})
versus the total decay width of the new scalar ($\Gamma_{\eta}$ )
where the palettes show the ranges of $m_{\beta}$ and $\Gamma_{\eta}/m_{\eta}$
respectively.

From (\ref{eq:muXX}), the signal strength has the value $c_{\beta}^{2}$
if the Higgs decaies only to SM particles, i.e., $\mathcal{B}_{BSM}=0$,
which makes small mixing values the prefered ones as shown in Fig.~\ref{fig:muxx}-left.
From Fig.~\ref{fig:muxx}-right, one notices that for $m_{\eta}\gtrsim30\,GeV$,
the total decay width has values $\Gamma_{\eta}\gtrsim1\,MeV$, and
it gets larger with large $m_{\eta}$ values. In Fig.~\ref{fig:DM},
we show the DM-nucleon direct detection cross section (\ref{eq:DD})
versus the DM mass (left) and the relative contribution of each annihilation
channel to the total annhilation cross section at the freeze--out
temperature (right).

From Fig.~\ref{fig:DM}-left, one remarks that viable dark vector
DM scenario is possible at all DM masses range, and future experiements
such as $Xenon-\,nT$~\cite{Aprile:2015uzo}, can probe a significant
part of the parameters space. Indeed, some of the benchmark points
can not be probed since they are below the neutrino floor. The freeze-out
parameter $x_{f}=m_{A}/T_{f}$, that is shown in the palette reads
typical values $x_{f}\sim17-28$. In order to figure out which the
DM annhiliation channels is (are) effiecient, we present in Fig.~\ref{fig:DM}-right
the relative contribution of each DM annihilation channel ($f\bar{f}$,
$WW$, $ZZ$, $hh$, $\eta\eta$ and $h\eta$) to the total thermally
averaged annihilation cross section at the freeze-out temperature
$T_{f}=m_{A}/x_{f}$ versus the DM mass.

Clearly from the Fig.~\ref{fig:DM}-right, one notes that the cross
section tends to be dominated by the annihilation into scalar channels
($hh$, $\eta\eta$ and $h\eta$) which are mediated by the of exchanging
$h$, $\eta$ and $A_{i}$ bosons for all values of the DM mass. In
addition the co-anihilation contribution (i.e., the second term in
(\ref{eff})) could be large as $20\%$ of the total thermally averaged
annihilation cross section. The fact that the scalar channels $\eta\eta$
is also dominant for light DM ($\mathcal{O}(GeV)$) means that light
DM implies light new scalr $\eta$. One has to mention aslo that the
annihilation into gauge bosons could be efficient for DM masses around
the $Z$ mass ($m_{Z}$) and also around $TeV$.

\begin{figure}[h!]
\begin{centering}
\includegraphics[width=0.48\textwidth]{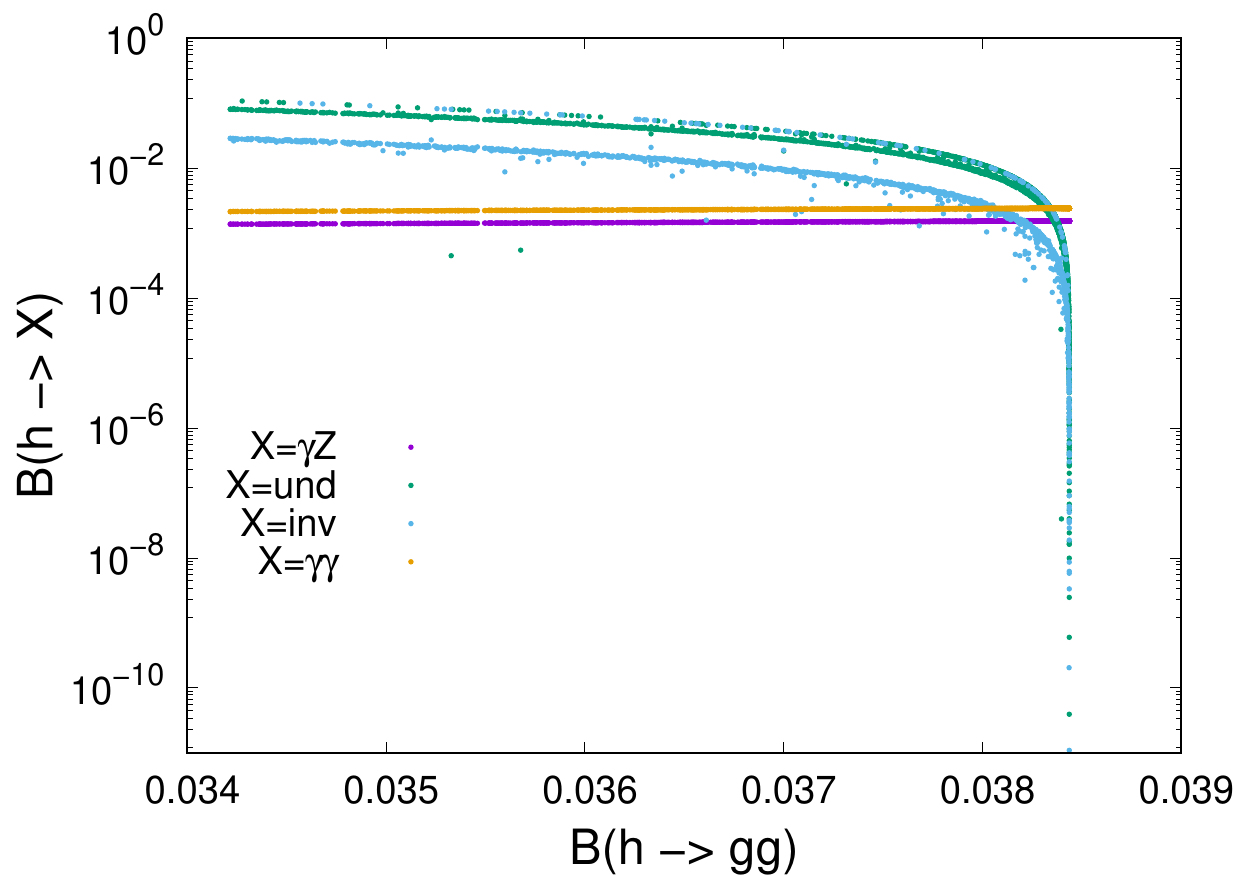}~\includegraphics[width=0.48\textwidth]{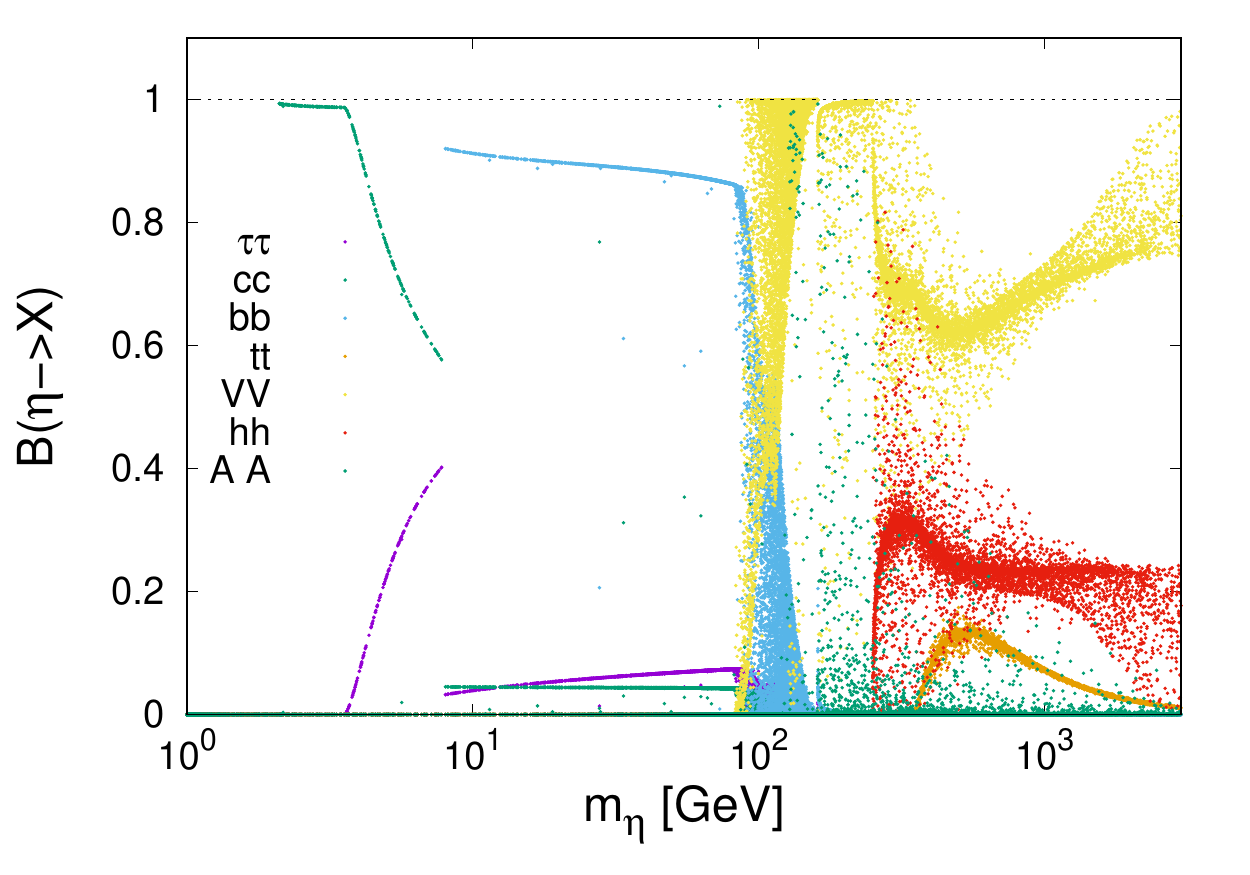} 
\par\end{centering}
\caption{Left: the Higgs branching ratios. Right: the branching ratios of the
new scalar decay versus its mass.}
\label{fig:decay} 
\end{figure}

In Fig.~\ref{fig:decay}, we present some branching ratios of the
Higgs (left), and the main branching ratios of the scalar $\eta$
(right), especially $\gamma\gamma,A_{i}A_{i},hh,t\bar{t},b\bar{b},VV$
as functions of $\eta$ mass.

One notices from Fig.~\ref{fig:decay}-left, that the branching ratios
of $h\to gg$, $h\to\gamma\gamma$ and $h\to\gamma Z$ have values
close the SM ones. Concerning the invisible and undetermined barnching
ratios ($h\to AA$ and $h\to\eta\eta$), they can be significant withing
the allowed range for some of the parameters space. From Fig.~\ref{fig:decay}-right,
one can learn that scalar $\eta$ decay can be dominated by a specific
contribution for some $m_{\eta}$ intervals. For instance, the decay
into light quarks, mainly $\eta\rightarrow c\bar{c}$, dominates for
the mass window $m_{\eta}\apprle8\,GeV$, while, it is dominated by
$\eta\rightarrow b\bar{b}$ for $8\,GeV\apprle m_{\eta}\apprle80\,GeV$.
For the mass window $80\,GeV\apprle m_{\eta}\apprle350\,GeV$ the
decay will be dominated by $\eta\rightarrow WW,ZZ$ as well for some
benchmark points with large scalar mass $m_{\eta}>2.5\,TeV$. The
branching ratios of $\eta\rightarrow hh$ adn $\eta\rightarrow t\bar{t}$
have maximal values of $0.35$ and $0.18$, respectivley for scalar
masses larger than $400\,GeV$. However, the invisible channel $\eta\rightarrow AA$
could be important for few benchmark points with mass betwenn $100\,GeV$
and $350\,GeV$.

\begin{figure}[h!]
\begin{centering}
\includegraphics[width=0.33\textwidth]{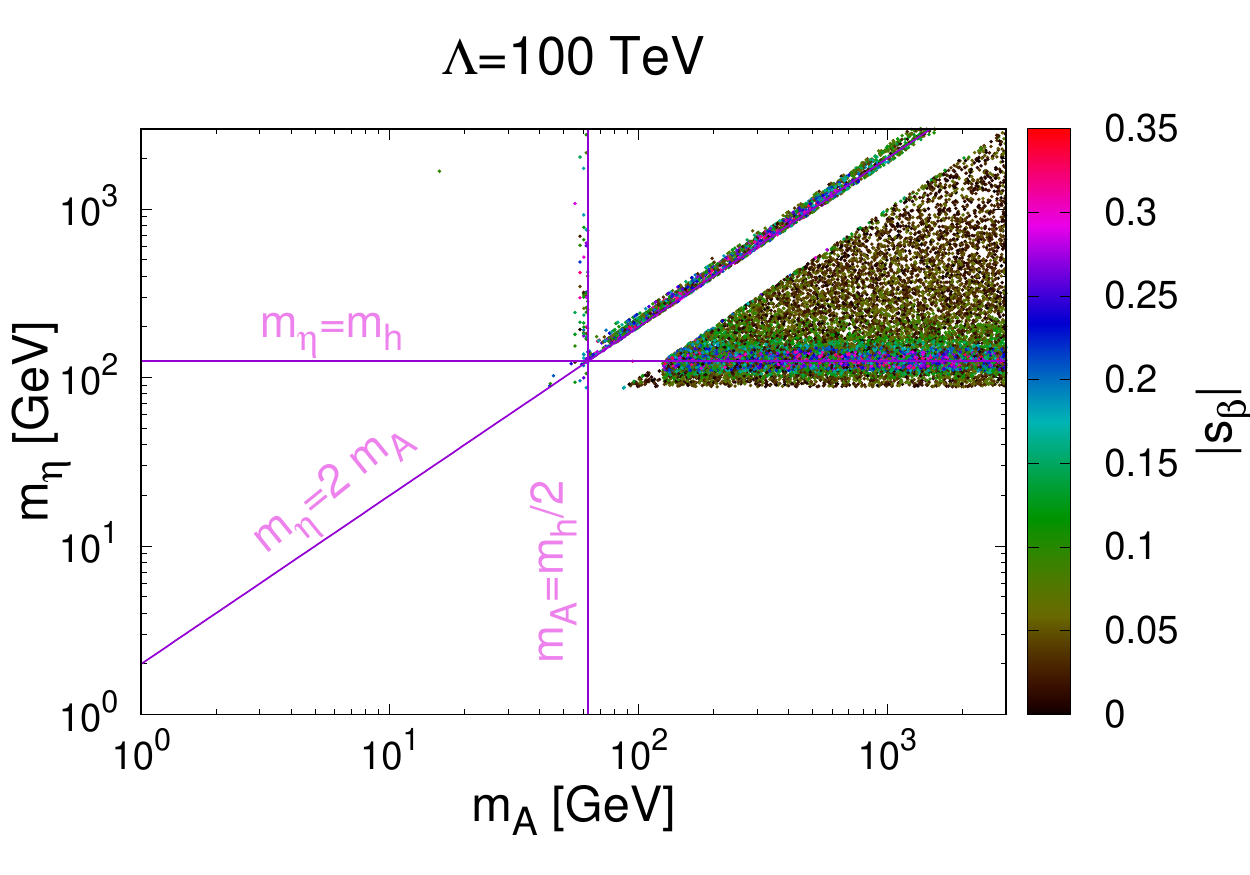}~\includegraphics[width=0.33\textwidth]{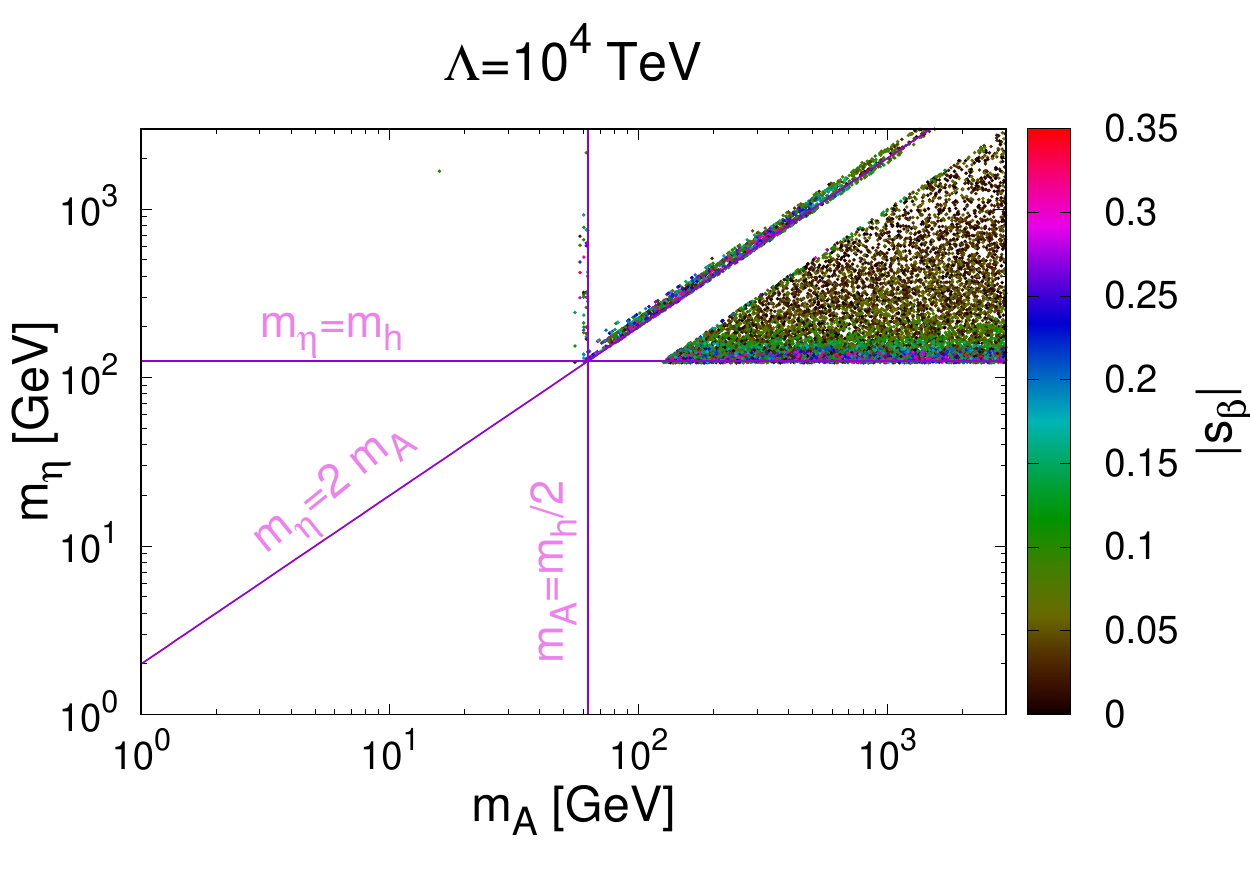}~\includegraphics[width=0.33\textwidth]{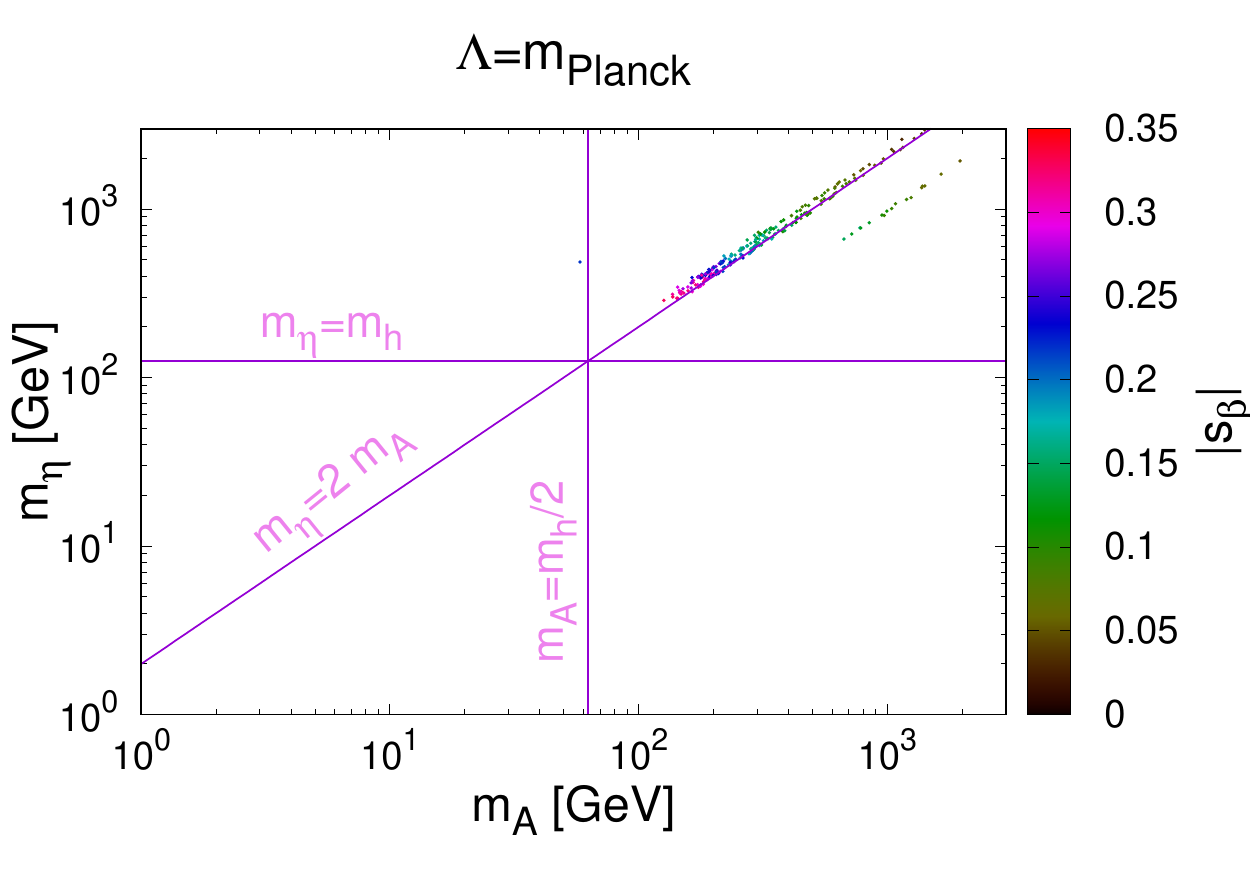}\\
 \includegraphics[width=0.33\textwidth]{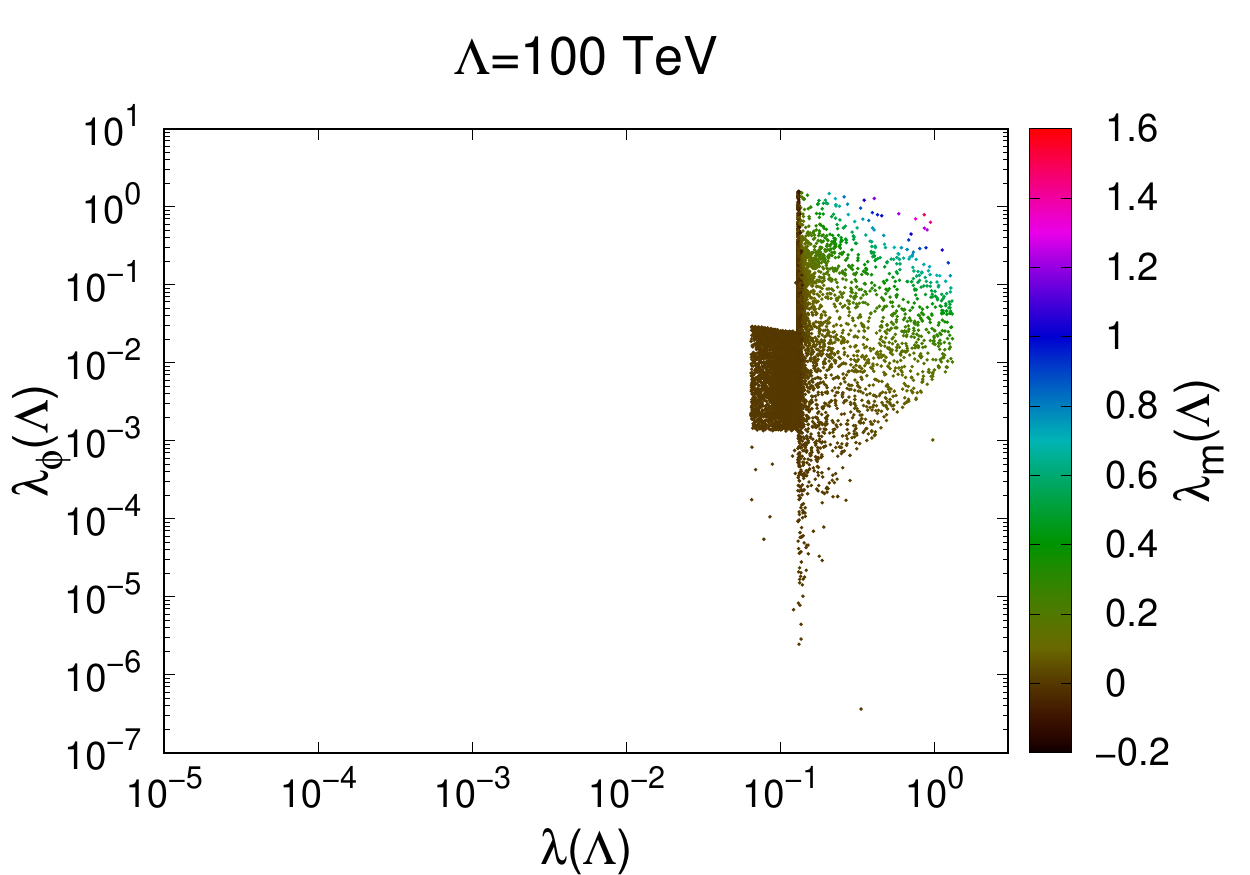}~\includegraphics[width=0.33\textwidth]{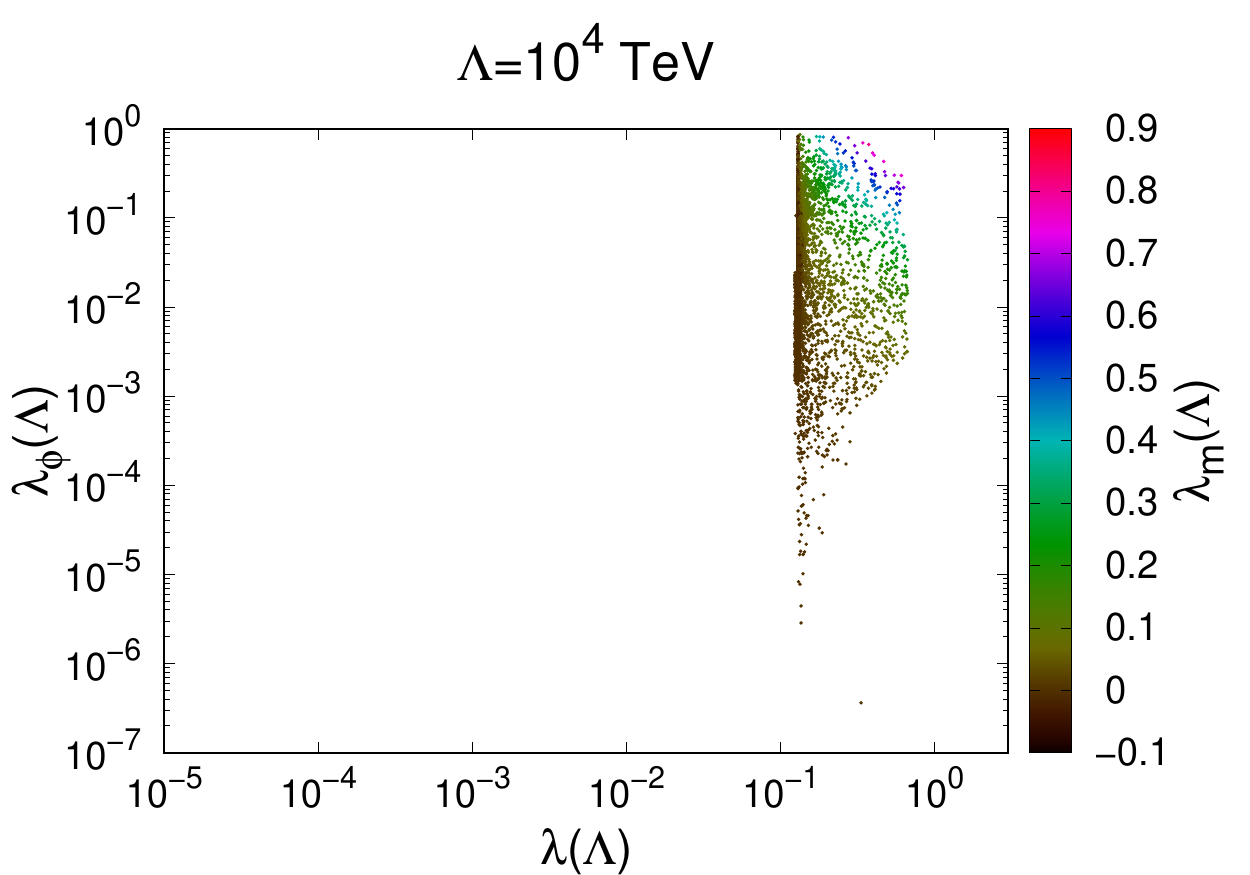}~\includegraphics[width=0.33\textwidth]{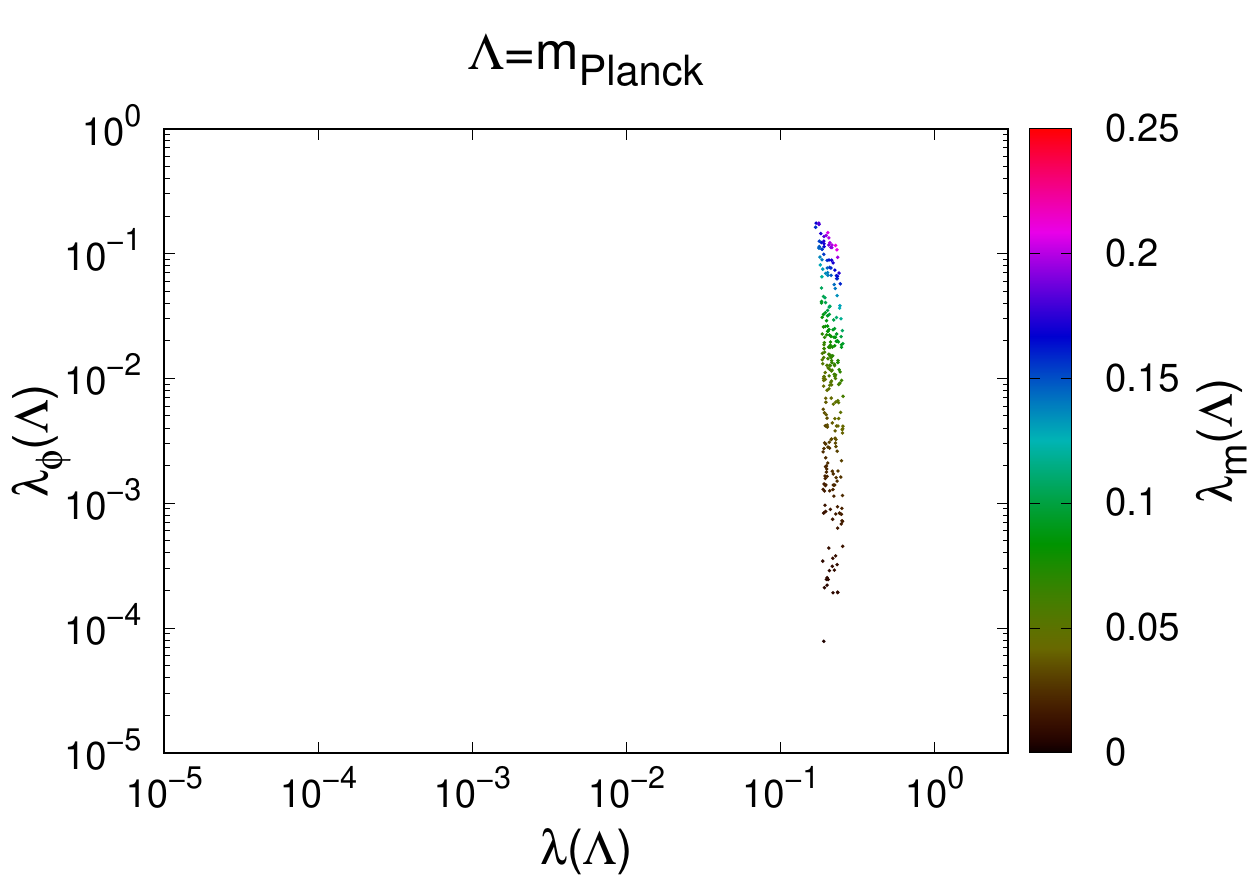}\\
 \includegraphics[width=0.33\textwidth]{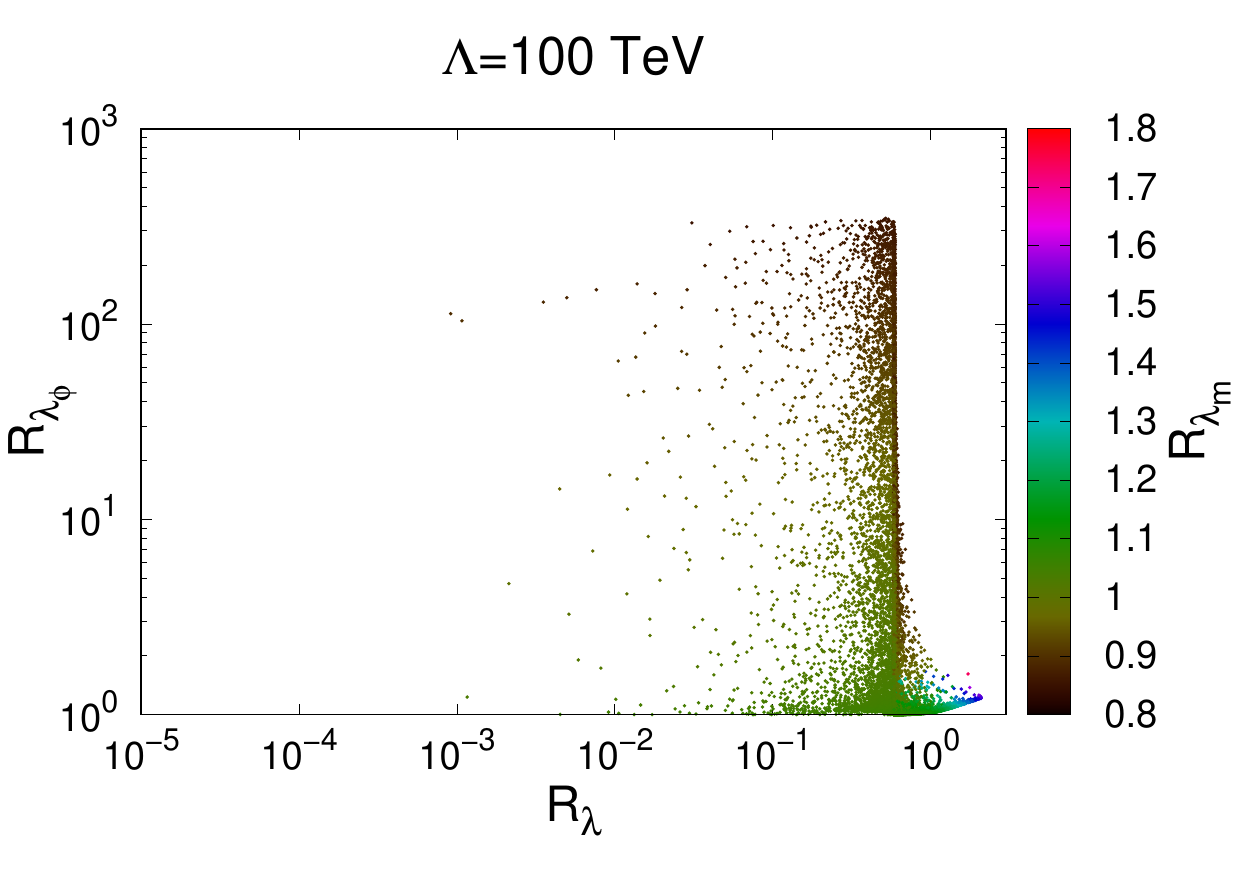}~\includegraphics[width=0.33\textwidth]{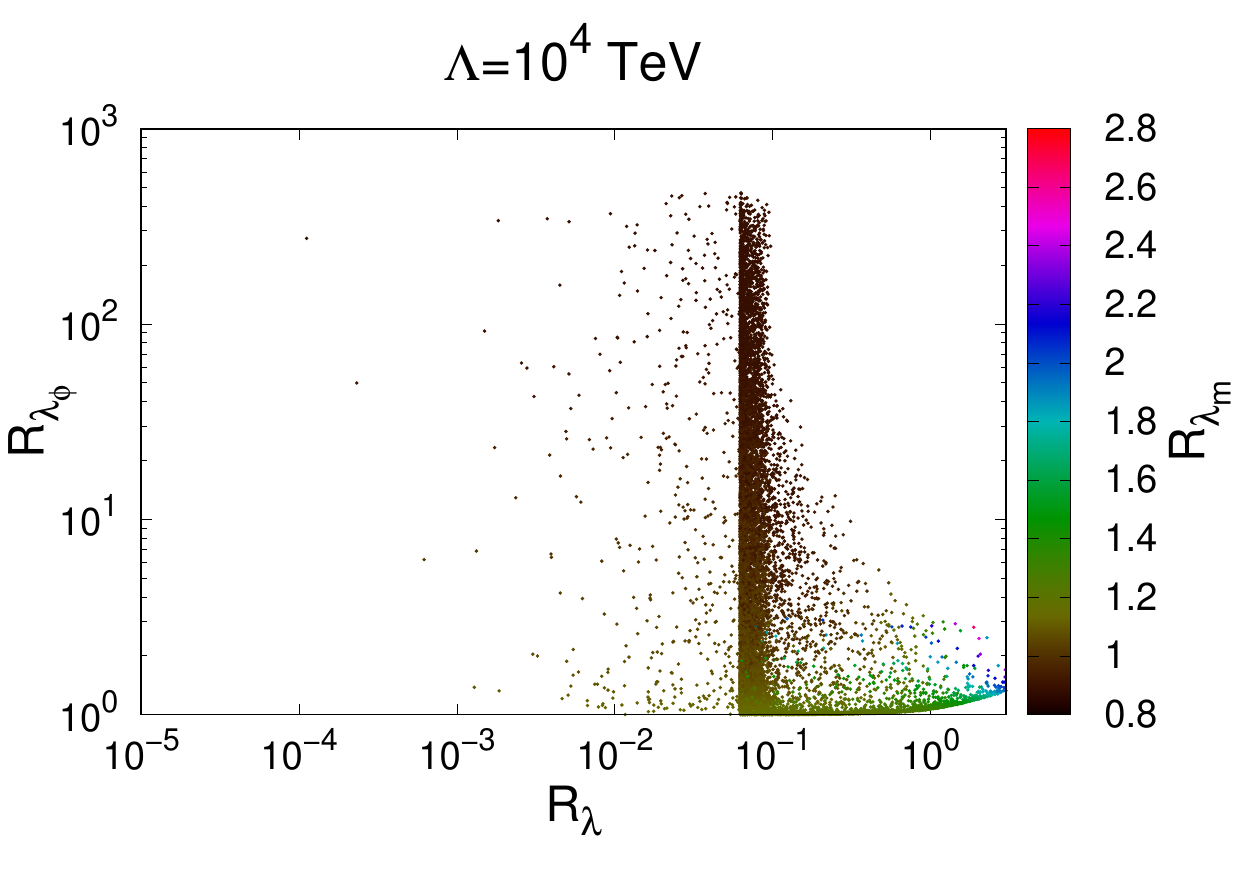}~\includegraphics[width=0.33\textwidth]{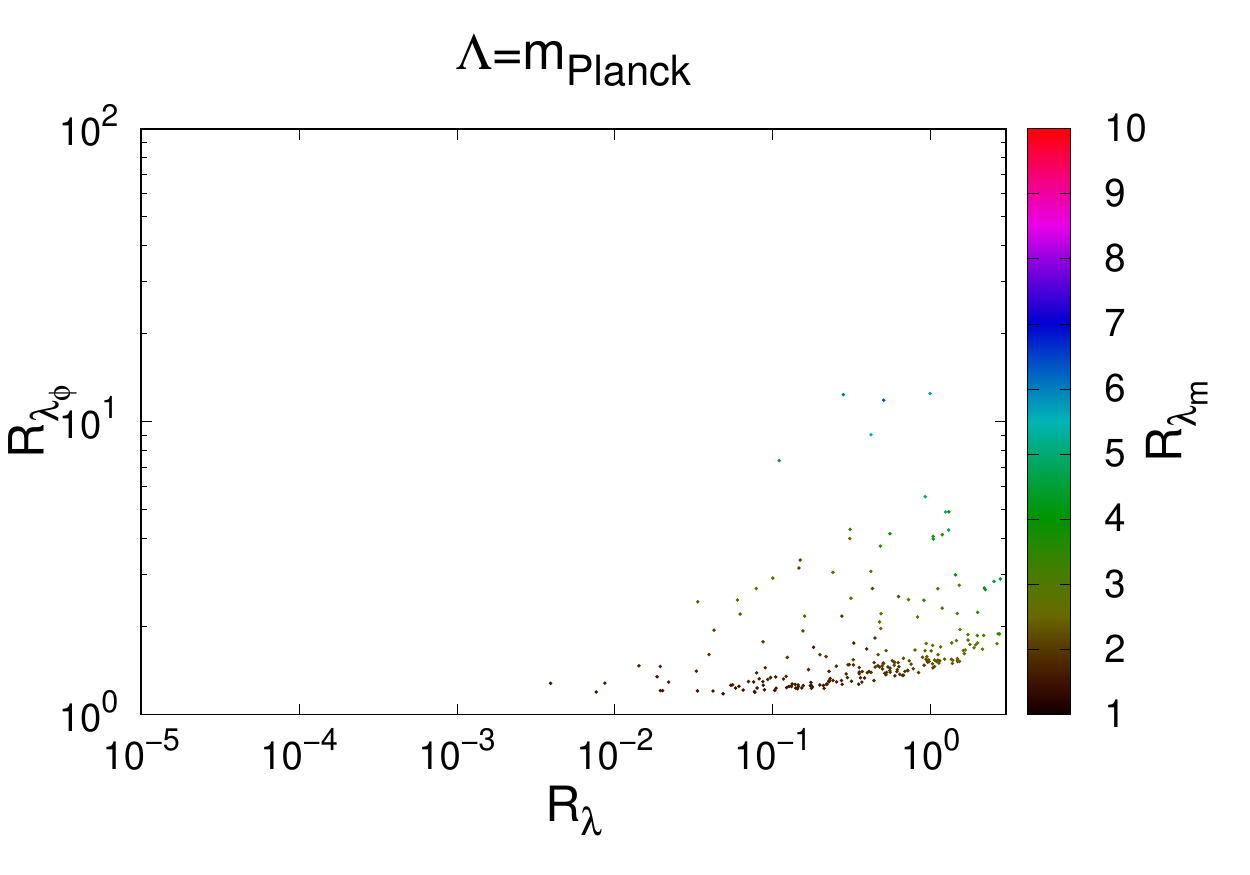} 
\par\end{centering}
\caption{Up: the updated version of Fig.~\ref{fig:SP}-top-left after considering
the abovementioned conditions at scales $\Lambda=100~TeV$ (left),
$\Lambda=10^{4}~TeV$ (middle) and $\Lambda=m_{Planck}$ (right).
Middle: the updated version of Fig.~\ref{fig:SP}-bottom-right within
similar assumptions. Bottom: we show the coupling values at the runnuing
scale, and in the bottom ones, we show their relativle enhancement
$R_{\lambda_{i}}(\Lambda)=\lambda_{i}(\Lambda)/\lambda_{i}(m_{Z})$.}
\label{PS} 
\end{figure}

By running the RGE in (\ref{RGE}) up to the high energy scales $\Lambda=100\,TeV,\,10^{4}\,TeV$
and $\Lambda=m_{Planck}$, we obtain the running diemensionless scalar
couplings. By imposing the conditions of perturbativity, vacuum stability
and unitarity at these energy scales, the parameters space get reduced as shown
in Fig.~\ref{PS}, where the plots top-left and bottom-right in Fig.~\ref{fig:SP},
are obtained after after considering the above mentioned conditions.

At higher scales, only benchmarks with non-suppressed $\lambda$ fulfills
the above-mentioned conditions. In addition, only benchmarks points
with positive $\lambda_{m}$ are favored. The enhancement in $\lambda_{\phi}$
could be two orders of magnitude lragere, which allows benchmark points
with small $\lambda_{\phi}$ values. For intance, for 20k bencmark
points shown in Fig.~\ref{fig:SP}, the conditions of perturbativity,
vacuum stability and unitarity at the scales $\Lambda=100\,TeV,\,10^{4}\,TeV$
and $\Lambda=m_{Planck}$, are filfilled only for $16700$, $11900$
and $260$, respectively.

\subsection{Triple Higgs Coupling}

In the SM, electroweak symmetry breaking relies on the parameters
in the Higgs scalar potential, namely on the choice $\mu^{2}<0$ and
$\lambda>0$. Hence, the partial experimental reconstruction of the
Higgs scalar potential through measuring the triple Higgs coupling,
$\lambda_{hhh}$, turns to be crucial to verify that the symmetry
breaking is due to a SM-like Higgs sector~\cite{Dolan:2012rv}. Not
only this, but it can also shed light on new physics~\cite{Efrati:2014uta}
knowing that in many extensions of the SM, $\lambda_{hhh}$ can be
modified by Higgs mixing effects or higher order corrections induced
by new particles as we will show below. Consequently, the measurement
of $\lambda_{hhh}$ is a crucial task in the LHC, although being challenging
~\cite{Baglio:2012np,Dolan:2012rv}, and future collider experiments.

Indirect probe of the triple Higgs coupling can be carried out through
investigating the loop effects in some observables such as the single
Higgs production~\cite{McCullough:2013rea,Gorbahn:2016uoy,Maltoni:2017ims},
and the electroweak precision observables~\cite{Kribs:2017znd}.
Using the 80 fb$^{-1}$ of LHC Run-2 data, and upon the assumption
that new physics can affect only $\lambda_{hhh}$, ATLAS collaboration
set recently the bound $-1.5<\lambda_{hhh}/\lambda_{hhh}^{SM}<6.7$
at $95\%$ CL~\cite{ATLAS:2021jki}. It should be noted that, direct
measurement of the triple Higgs coupling at the LHC is possible and
can be achieved through the di-Higgs production. This production is
dominated by the gluon-gluon fusion process. In the SM, the production
has two main contributions originating from the triangle diagram induced
by the triple Higgs coupling, and from the box diagram with the top
quark running in the loop. As noted in~\cite{Chen:2019fhs}, the
two amplitudes, corresponding to the two contributions, interfere
destructively. Consequently, at next-to-next-to-next-to-leading order
(N$^{3}$LO) and after including finite top quark mass effects, the
estimated cross section at $14\,TeV$ LHC turn to be small and equal
to $38.65\,fb$~\cite{Chen:2019fhs}.

The analysis of the potential of measuring the di-Higgs production
in the decay channels $b\bar{b}b\bar{b}$, $b\bar{b}\tau^{+}\tau^{-}$,
$b\bar{b}WW^{*}$, $\gamma\gamma b\bar{b}$, $\gamma\gamma WW^{*}$
and $WW^{*}WW^{*}$, has been carried out in Refs.~\cite{Dolan:2012rv,Baglio:2012np,Barr:2013tda,Li:2013flc,Cao:2015oaa,Li:2015yia,Cao:2015oxx,Cao:2016zob,He:2015spf,Huang:2017jws,Lu:2015jza,Chang:2018uwu,Papaefstathiou:2012qe,Kim:2018cxf,Buchalla:2018yce,Kim:2019wns,Li:2019uyy}.
At $13\,TeV$ LHC with the luminosity of $36.1\,fb^{-1}$, the combination
of the six analyses results in the constraint $-5<\lambda_{hhh}/\lambda_{hhh}^{SM}<12$
at $95\%$ CL~\cite{Abdughani:2020xfo}. It is expected that, the
sensitivity will be highly improved at the high-luminosity upgrade
of the LHC (HL-LHC)~\cite{Cepeda:2019klc} and future hadron colliders~\cite{Contino:2016spe}.
For instances, the future circular hadron collider, FCC-$hh$, with
a center-of-mass energy of $100\,TeV$ and an integrated luminosity
of $30\,ab^{-1}$ of data will allow reaching a $5\%$ accuracy (at
$68\%$ CL) on the measurement of the triple Higgs coupling~\cite{Goncalves:2018qas,Chang:2018uwu,Cepeda:2019klc}.

The initial phase of the International Linear Collider (ILC) with
a center-of-mass energy of $\sqrt{s}=$ $250\,GeV$ cannot directly
probe $\lambda_{hhh}$ via di-Higgs production~\cite{Fujii:2017vwa}.
However, this is not the case regarding single-Higgs production where
an analysis using $2\,ab^{-1}$ of data can allow a measurement to
$49\,\%$ accuracy, at $68\,\%$ CL~\cite{deBlas:2019rxi}. Higher
precision of $27\,\%$ or $10\,\%$ , at $68\,\%$ CL, could possibly
be reached using the data from ILC extensions to $500\,GeV$ ($4\,ab^{-1}$)
or $1\,TeV$ ($8\,ab^{-1}$) respectively~\cite{Fujii:2015jha,Braathen:2019zoh}.
Moreover, at the same value of the confidence level, a more higher
accuracy of $0.93<\lambda_{hhh}/\lambda_{hhh}^{SM}<1.11$ is possible
to be reached using the combination of $1\,ab^{-1}$ of data at $380\,GeV$,
$2.5\,ab^{-1}$ at $1.5\,TeV$, and $5\,ab^{-1}$ at $3\,TeV$ expected
to be collected in the CLIC project~\cite{Abramowicz:2016zbo,Charles:2018vfv,Roloff:2019crr,Braathen:2019zoh}.

In our model, the triple Higgs coupling $\lambda_{hhh}$ gets modified
due to the mixing with the scalar $\eta$, and in addition, it receives
new one-loop contributions by the scalar $\eta$ and the new gauge
bosons $A_{i}$. Taking these contributions into account, we can parameterize
$\lambda_{hhh}$ as\textbf{ 
\begin{equation}
\lambda_{hhh}=\lambda_{hhh}^{SM}(1+\Delta_{hhh}),
\end{equation}
}where the\textbf{ }one-loop triple Higgs coupling in the SM is given
by~\cite{Kanemura:2002vm,Kanemura:2004mg} 
\begin{equation}
\lambda_{hhh}^{SM}\simeq\frac{3m_{h}^{2}}{\upsilon}\left[1-\frac{m_{t}^{4}}{\pi^{2}\upsilon^{2}m_{h}^{2}}\right].\label{eq:SMLhhh}
\end{equation}

According to the ILC physics group, the triple Higgs coupling can
be measured at $\sqrt{s}=500\,GeV$ within the integrated luminosity
$\mathcal{L}=500\,fb^{-1}$ with an accuracy less or equal to $20\,\%$~\cite{Fujii:2015jha}.
This implies that for the parameters space where $m_{\eta}\gg m_{h}$
\footnote{In this model, the relevant process that probes the triple Higgs coupling
at the ILC ($e^{+}e^{-}\rightarrow Zhh\,@\,500\,GeV$), occurs via
another Feynmann diagram mediated by the new scalar $\eta$. Therefore,
if $\text{\ensuremath{m_{\eta}}}\gtrsim400\,GeV\gg m_{h}$ this new
daigram is subleading and therefore the triple Higgs coupling can
be probed similarly to the case of single Higgs models.}, our model can tested at future linear colliders. Here, we estiamte
the parameter $\Delta_{hhh}$ following~\cite{Ahriche:2013vqa},
where the Higgs trilinear self-coupling can be considered as the third
derivative of the Higgs one-loop effective potential 
\begin{equation}
\lambda_{hhh}=\frac{\partial^{3}V_{eff}}{\partial h^{3}},\label{eq:Lhhh}
\end{equation}
where, the zero temperature one-loop effective potential $V_{eff}(h',\eta')$
is described in Appendix~\ref{sec:One-Loop}. Therefore, one writes 
\begin{equation}
\begin{array}{c}
\lambda_{hhh}=\left.c_{\beta}^{3}\,\frac{\partial^{3}V_{eff}}{\partial h'^{3}}-3c_{\beta}^{2}\,s_{\beta}\,\frac{\partial^{3}V_{eff}}{\partial\eta'\partial h'^{2}}+3c_{\beta}\,s_{\beta}^{2}\,\frac{\partial^{3}V_{eff}}{\partial\eta'^{2}\partial h'}-s_{\beta}^{3}\frac{\partial^{3}V_{eff}}{\partial\eta'^{3}}\right|_{h'=\eta'=0}.\end{array}\label{Lhhh}
\end{equation}

\begin{figure}[h!]
\begin{centering}
\includegraphics[width=0.5\textwidth]{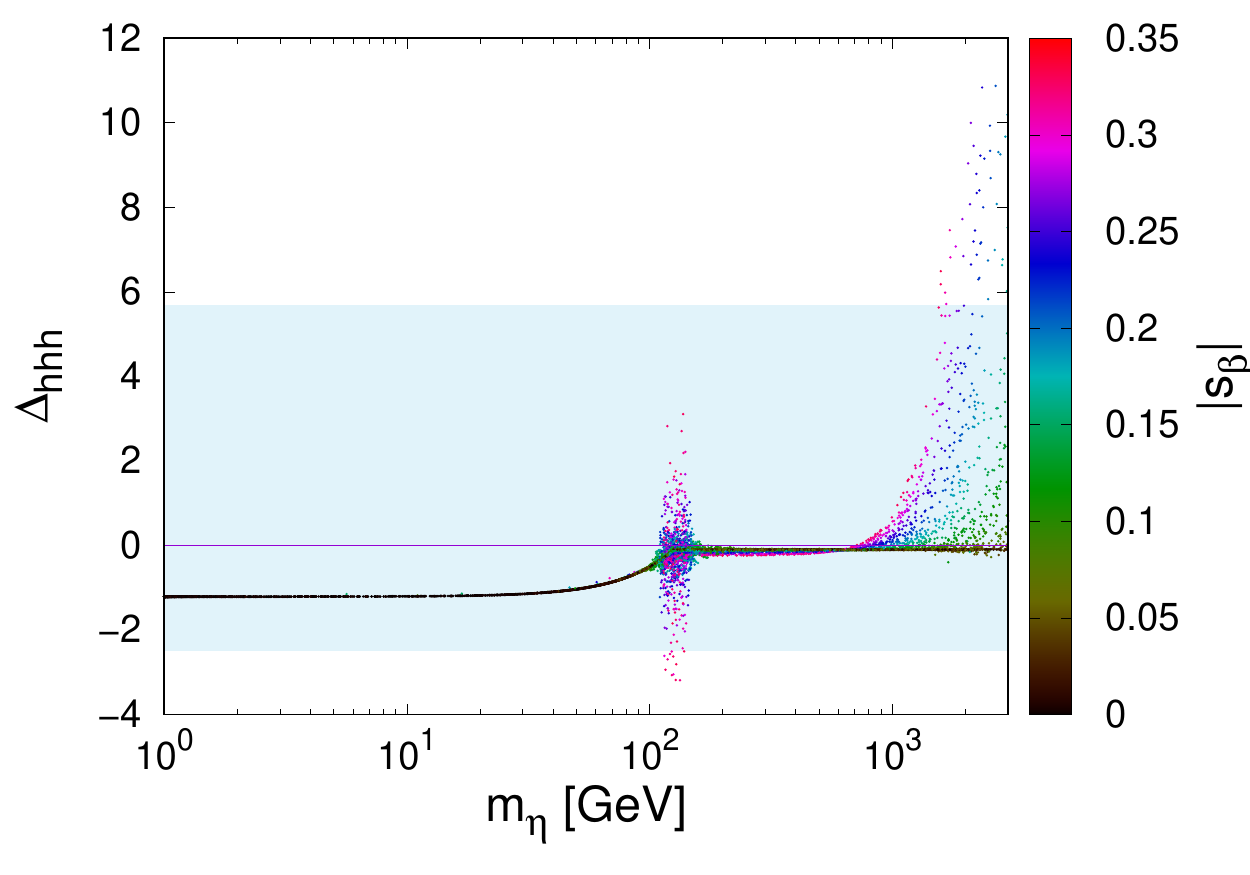} 
\par\end{centering}
\caption{The triple Higgs relative enhancement $\Delta_{hhh}$ versus the new
scalar mass, where the palette shows the scalar mixing. The skyblue
band represents the allowed values of $\Delta_{hhh}$ by the recent
ATLAS measurments~\cite{ATLAS:2021jki}.}
\label{fig:D} 
\end{figure}

At tree-level, these scalar triple couplings correspond to the $\rho_{h}$
parameter given in (\ref{eq:rho}). For the bemchmarks points used
previously, we show the Higgs triple coupling relative enhancement
in Fig.~\ref{fig:D}.

The effect of these extra contributions can be either constructive
or destructive according to the $\eta$ mass range. For instance,
from Fig.~\ref{fig:hh}, one notices that for scalar mass $m_{\eta}<m_{h}$
the coupling (\ref{eq:Lhhh}) is almsot supressed. While for the degenerate
case $m_{\eta}\sim m_{h}$, we could have either enhancment or suppression.
For $\eta$ mass larger than $1.2\,TeV$, some benchgmark points are
already supressed by the ATLAS recent measurments~\cite{ATLAS:2021jki}.

\subsection{The Di-Higgs Production}

\begin{figure}
\includegraphics[width=0.7\textwidth]{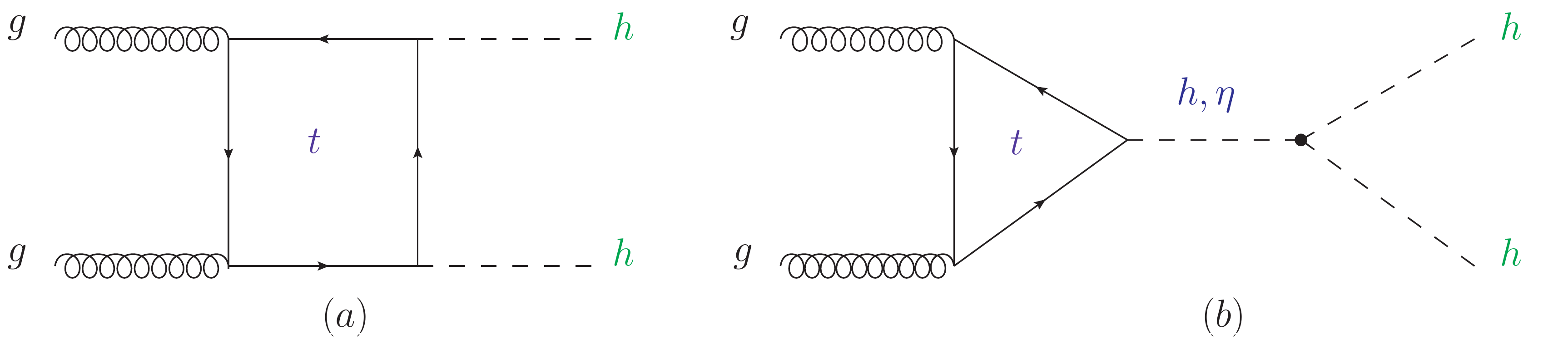}\caption{Feynman diagrams that contribute to the di-Higgs production via gluon
fusion. The left (right) diagram is referred to be the box (triangle)
in the literature. }
\label{fighh} 
\end{figure}

The di-Higgs production is not only interesting as its measurement
allows to determine the trilinear Higgs couplings but also to describe
the EWSB, i.e., it occurs via one Higgs or more. The triple Higgs
coupling can be measured directly in di-Higgs boson production at
$ILC$ through double Higgs-strahlung off $W$ or $Z$ bosons~\cite{Gounaris:1979px,Barger:1988jk,Ilyin:1995iy,Djouadi:1999gv,Tian:2010np}
, $WW$ or $ZZ$ fusion~\cite{Ilyin:1995iy,Djouadi:1999gv,Tian:2010np,Boudjema:1995cb,Barger:1988kb,Dobrovolskaya:1990kx,Abbasabadi:1988ja}
, also through gluon-gluon fusion in $pp$ colisions at the LHC~\cite{Glover:1987nx,Plehn:1996wb,Dawson:1998py}.
The Higgs pair production processes in this model can be achieved
three via Feynman diagrams as shown in Fig.~\ref{fighh}. Indeed,
there are two triangle diagrams mediated by the Higgs field $h$ and
the new singlet scalar field $\eta$ instead of one diagram in the
SM.

\begin{figure}[h!]
\begin{centering}
\includegraphics[width=0.48\textwidth]{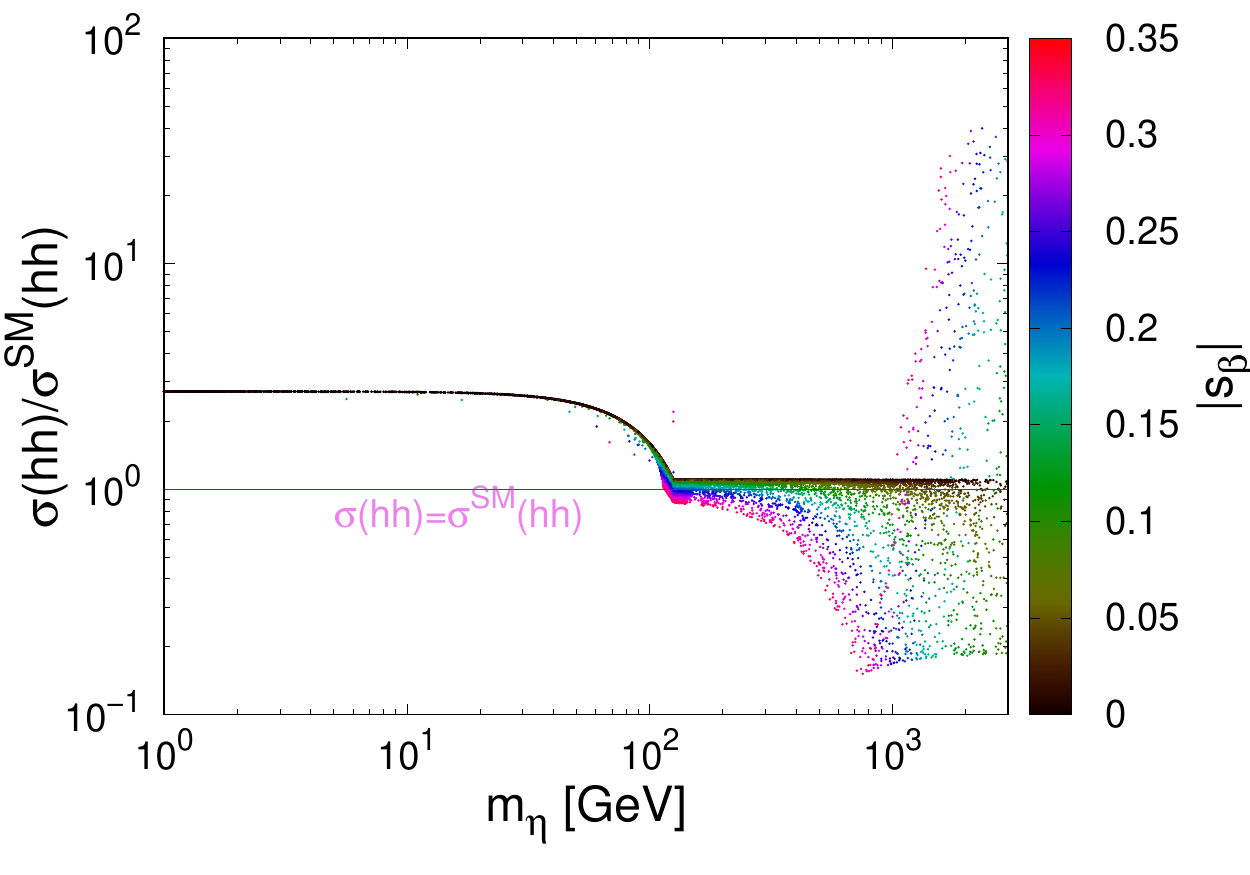}~\includegraphics[width=0.48\textwidth]{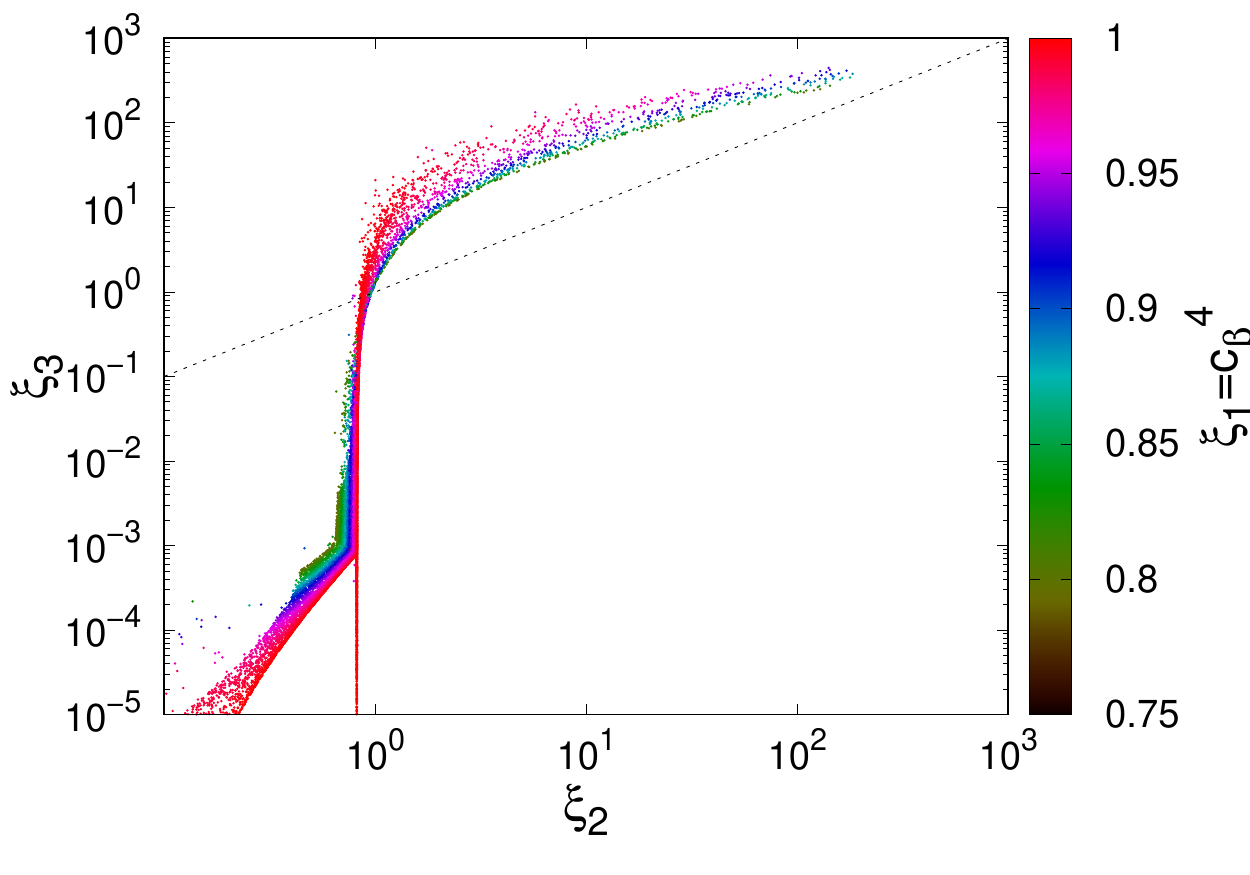} 
\par\end{centering}
\caption{Left: the di-Higgs production cross section at $14\,TeV$, scaled
by the SM value, where the palette shows the scalar mixing. Right:
the parameters $\xi$'s (\ref{eq:coef}) for the benchmark points
used previously. The dashed line represent $\xi_{2}=\xi_{3}$, above which the destructive intereference contirubtion in (\ref{eq:cro})
is larger, and hence the total crosss section is reduced. }
\label{fig:hh} 
\end{figure}

In the SM, the di-Higgs production cross section has three contributions
\begin{equation}
\sigma^{SM}\left(hh\right)=\sigma_{\Square}+\sigma_{\triangle}+\sigma_{\triangle\Square},
\end{equation}
which correspond to the box ($\sigma_{\Square}=70.1\,fb$), triangle
($\sigma_{\triangle}=9.66\,fb$); and interference ($\sigma_{\triangle\Square}=-49.9\,fb$),
respectively~\cite{Spira:1995mt}. In this model, the di-Higgs production
cross section can be written as 
\begin{equation}
\sigma\left(hh\right)=\xi_{1}\sigma_{\Square}+\xi_{2}\sigma_{\triangle}+\xi_{3}\sigma_{\triangle\Square},\label{eq:cro}
\end{equation}
where the SM corresponds to $s_{\beta}=0$, i.e., $\xi_{1}=\xi_{2}=\xi_{3}=1$.
The coeffetions $\xi_{i}$ in our model are modified with respect
to the SM as 
\begin{align}
\xi_{1}=c_{\beta}^{4},\,\xi_{2} & =\left(c_{\beta}\frac{\rho_{h}}{\lambda_{hhh}^{SM}}+s_{\beta}\frac{\rho_{2}}{\lambda_{hhh}^{SM}}\frac{s-m_{h}^{2}}{s-m_{\eta}^{2}}\right)^{2},\,\xi_{3}=c_{\beta}^{2}\left(c_{\beta}\frac{\rho_{h}}{\lambda_{hhh}^{SM}}+s_{\beta}\frac{\rho_{2}}{\lambda_{hhh}^{SM}}\frac{s-m_{h}^{2}}{s-m_{\eta}^{2}}\right),\label{eq:coef}
\end{align}
with $\rho_{h}$ and $\rho_{2}$ are defined in (\ref{eq:rho}), $\lambda_{hhh}^{SM}$
the Higgs triple coupling in the SM is given in (\ref{eq:SMLhhh});
and $\sqrt{s}$ is the CM collision energy, which we will consider
to be $\sqrt{s}=14\,\mathrm{TeV}$. In Fig.~\ref{fig:hh}-left, we
show the di-Higgs production cross section (\ref{eq:cro}) at LHC14
scaled by its SM value versus the new scalar mass $m_{\eta}$ and
the scalar mixing (in the palette), and in Fig.~\ref{fig:hh}-right,
we present the parameters $\xi$'s (\ref{eq:coef}) for the benchmark
points used previously.

Clearly for lighter new scalar $m_{\eta}<m_{h}$ the di-Higgs cross
section masses is larger than the SM values by around 180 \%, while,
for heavier $\eta$ scalar $m_{\eta}>m_{h}$, it lies beween -92 \%
and 3000 \%. Indeed, not all of the benchmark points would be in agreement
with current data. In order to understand these results, Fig.~\ref{fig:hh}-right
shows the relation beween the patameters $\xi_{2}$ and $\xi_{3}$,
where $\xi_{1}$ is shown in the palette. Since the interference term
in (\ref{eq:cro}) is negative, then the benchmark points above the
straight line $\xi_{2}=\xi_{3}$ in Fig.~\ref{fig:hh}-right correspond
to larger cross section values. For lighter $\eta$ scalar $m_{\eta}<m_{h}/2$,
where the undetermined Higgs decay channel $h\rightarrow\eta\eta$
is open, more negative searches can be used to constrain the parameters
space, especially via the signatures $h\rightarrow\eta\eta\rightarrow bb\mu\mu$~\cite{ATLAS:2021ypo},
$h\rightarrow\eta\eta\rightarrow4b$~\cite{Aad:2020rtv}; and $h\rightarrow\eta\eta\rightarrow\gamma\gamma jj$~\cite{Aaboud:2018gmx}.
This analysis is postponed for a future work~\cite{next}.

\section{Conclusion\label{sec:Conc}}

In this work, we have considered an extension of the SM by enlarging
the SM gauge symmetry by a non-abelian gauge group $SU(2)_{HS}$.
In this case, the scalar sector of the model contains an extra new
Higgs doublet that is required to spontaneously break the $SU(2)_{HS}$
gauge symmetry. The mixing of the extra doublet with the SM Higgs
results, in the mass eigenstates basis, in two scalars denoted by
$h$ and $\eta$. In our analysis, we identified $h$ as the SM neutral
Higgs-like boson and adopted its mass to be $125\,GeV$. An important
feature of the model is that the new gauge bosons, $A_{i}^{\mu}$
with i=1,2,3, associated with the group $SU(2)_{HS}$ are exactley
degenerate in mass and can serve as vector DM candidates that interact
with the SM through the Higgs portal.

To investigate such a possibility for DM candidate, we first considered
all relevant constraints on the model, which include both theoretical
and experimental ones, such as perturbative unitarity, vacuum stability,
perturbativity, experimental bound on the DM direct detection, the
observed DM relic density, the constraints from the Higgs decay where
the invisible or/and undetermined branching ratio where the Higgs
total decay width must respect the existing experimental constraints.
As a result, we showed that it is possible to have viable parameters
space in which the masses of the DM candidates lies from few GeV to
the TeV scale, which is still within the reach of high energy collider
experiments. Regarding the limits from direct detection DM experiements,
it is easily acomodated in this model for most of the values of DM
and new scalar masses. In adition, the observed relic density values
can be achieved by many anihilation channels acording to the DM mass,
and on top of that, the co-anihilation ($A_{i}A_{j}\rightarrow hA_{k}$)
effcect could be importan as it may reach 25 \% of the total thermally
averaged anihilation cross section.

In our wrok, we considered the conditions of perturbativity, vacuum
stabilty and perturbative unitarity, but these conditions may not
be fulfilled at higher scales. By runing the quartic scalar and gauge
couplings at higher scale $\Lambda$ using the RGE (\ref{RGE}), we
found that 16.5 \%, 40 \% and 98.7 \% of the benchmark points will
be ruled out due these conditions at scale values $\Lambda=100\,TeV,\,10^{4}\,TeV,\,m_{Planck}$,
respectively.

In the decoupling limit $m_{\eta}>>m_{h}$, as in many extensions
of the SM, the Higgs mixing effects and the presence of new fields
coupled to the Higgs doublet induces significant corrections to the
SM prediction of the triple Higgs self-couplings $\lambda_{hhh}=(1+\Delta_{hhh})\lambda_{hhh}^{SM}$.
We have found that, up to one-loop level, the effect of the new scalar
$\eta$ and the vector DM $A_{i}$ leads to a relative enhancement
($\Delta_{hhh}$) that lies between -250 \% and +1200 \%. Indeed,
part of these benchamrk points are already esxcluded by the recent
measerments by ATLAS~\cite{ATLAS:2021jki}. However, in the case
$m_{\eta}<m_{h}$, the cross the di-Higgs gets enhanced by around
180 \%, which makes the signatures $h\rightarrow\eta\eta\rightarrow bb\mu\mu$,
$h\rightarrow\eta\eta\rightarrow4b$ and $h\rightarrow\eta\eta\rightarrow\gamma\gamma jj$
very useful to put more constraints on the model free parameters.

\appendix

\section{The One-Loop Effective Potential\label{sec:One-Loop}}

The zero temperature one-loop effective potential can be given in
the $\overline{DR}$ scheme by

\begin{align*}
V_{eff}\left(h',\eta'\right) & =V^{0}\left(h',\eta'\right)+\frac{1}{64\pi^{2}}\sum_{i}n_{i}m_{i}^{4}\left(h',\eta'\right)\left(\log\left(\frac{m_{i}^{2}\left(h',\eta'\right)}{\Lambda^{2}}\right)-\frac{3}{2}\right),
\end{align*}
where, $V^{0}\left(h',\eta'\right)$ is the tree-level potential,
$n_{i}$ is the number of internal degrees of freedom of the $i\,$th
particle ($n_{h}=n_{\eta}=1$, $n_{Z}=3$, $n_{W}=6$, $n_{t}=-12$
and $n_{A_{i}}=9$). Here, $m_{i}^{2}\left(h',\eta'\right)$ are the
field-dependent squared masses; and $\Lambda$ is the renormalization
scale, which we will choose to be the Higgs mass $\Lambda=125.18\,\mathrm{GeV}$.

The field-dependent squared masses $m_{i}(h',\eta')$ of all the contributing
particles, so we have the field-dependent masses of the Goldstone
bosons $\chi$ and $\xi$

\begin{equation}
m_{\chi}^{2}=\mu^{2}+\lambda h'^2 +\frac{\lambda_{m}}{2}\eta'^{2},\,m_{\xi}^{2}=\mu_{\phi}^{2}+\lambda_{\phi}\eta'^{2}+\frac{\lambda_{m}}{2} h'^2 .
\end{equation}

The field-dependent masses of the electroweak gauge bosons and top
quark are given in the symmetric phase (i.e., for $<h>=<\phi>=0$) by

\begin{equation}
m_{t}^{2}=\frac{y_{t}^{2}}{2} h'^2 ,\,m_{W}^{2}=\frac{g_{2}^{2}}{4} h'^2 ,\,m_{W^{3}}^{2}=\frac{g_{2}^{2}}{4} h'^2 ,\,m_{W^{3}-B}^{2}=\frac{g_{1}^{2}}{4} h'^2 ,m_{B}^{2}=\frac{g_{1}^{2}}{4} h'^2 ,\,m_{A}^{2}=\frac{g_{\phi}^{2}}{4} \eta'^{2},
\end{equation}
where the diagonalization of the \{$W^{3}-B$\} matrix gives $m_{\gamma}=0$
and $m_{Z}^{2}=\frac{\left(g_{1}^{2}+g_{2}^{2}\right)}{4} h'^2 $.
Here, $y_{t}$ denotes the top-quark Yukawa coupling, and $g_{1}$,
$g_{2}$ and $g_{\phi}$ are the gauge couplings of $U(1)_{Y}$, $SU(2)_{L}$
and $SU(2)_{HS}$, respectively.

For the field-dependent masses of $h$ and $\eta$ can be obtained
as the eigensvalues of the squared mass matrice in the basis $\left\{ h',\eta'\right\} $,
which is given by

\begin{equation}
M^{2}=\left(\begin{array}{cc}
A & C\\
C & B
\end{array}\right),
\end{equation}
with $A=\mu^{2}+3\lambda h'^{2}+\frac{\lambda_{m}}{2}\eta'^{2},\,B=\mu_{\phi}^{2}+3\lambda_{\phi}\eta'^{2}+\frac{\lambda_{m}}{2} h'^{2},\,C=\lambda_{m} h' \eta'$.
Then, the field dependant eigenmasses are given by $m_{h,\eta}^{2}\left(h',\eta'\right)=\frac{1}{2}\left(A+B\mp\sqrt{(A-B)^{2}+C^{2}}\right)$.

\section{The Cross Section for $hh$, $\eta\eta$, $h\eta$, $Ah$ \& $A\eta$\label{sec:XSection}}

Here, we give the formulas of the parameters used in the cross section
of the final states $hh$, $\eta\eta$, $h\eta$, $Ah$ \& $A\eta$
given in (\ref{M2}), (\ref{MM2}) and (\ref{XS}). We denote by $H$
either $h$ or $\eta$, so the parameters are given by 
\begin{align}
A & =\frac{2m_{H}^{2}-s}{2m_{A}^{2}},\,B=\frac{s}{2m_{A}^{2}}\sqrt{\left(1-\frac{4m_{H}^{2}}{s}\right)\left(1-\frac{4m_{A}^{2}}{s}\right)},\,Q_{0}=\frac{\sqrt{1-\frac{4m_{H}^{2}}{s}}}{16\pi},\nonumber \\
Q_{1} & =\frac{2q^{4}}{9\upsilon_{\phi}^{4}m_{A}^{4}}(48m_{A}^{8}-32m_{A}^{6}m_{H}^{2}+24m_{A}^{4}m_{H}^{4}-16m_{A}^{4}m_{H}^{2}s+4m_{A}^{4}s^{2}-8m_{A}^{2}m_{H}^{6}+4m_{A}^{2}m_{H}^{4}s+m_{H}^{8}),\nonumber \\
Q_{2} & =\frac{4q^{4}}{9\upsilon_{\phi}^{4}m_{A}^{2}\left(s-2m_{H}^{2}\right)}(-48m_{A}^{8}+32m_{A}^{6}m_{H}^{2}-8m_{A}^{4}m_{H}^{4}+4m_{A}^{4}s^{2}-8m_{A}^{2}m_{H}^{6}+4m_{A}^{2}m_{H}^{2}s^{2}-2m_{A}^{2}s^{3}\nonumber \\
 & +3m_{H}^{8}-m_{H}^{4}s^{2})+\frac{4q^{2}\Re(R)}{9\upsilon_{\phi}^{4}m_{A}^{2}\left(s-2m_{H}^{2}\right)}(-48m_{A}^{6}m_{H}^{2}+24m_{A}^{6}s+16m_{A}^{4}m_{H}^{4}-4m_{A}^{4}s^{2}-4m_{A}^{2}m_{H}^{6}\nonumber \\
 & +10m_{A}^{2}m_{H}^{4}s-8m_{A}^{2}m_{H}^{2}s^{2}+2m_{A}^{2}s^{3}-2m_{H}^{6}s+m_{H}^{4}s^{2}),\nonumber \\
Q_{3} & =\frac{4q^{4}}{9\upsilon_{\phi}^{4}}(-4m_{A}^{4}+4m_{A}^{2}m_{H}^{2}-6m_{A}^{2}s+m_{H}^{4}+2m_{H}^{2}s)+\frac{8q^{2}\Re(R)}{9\upsilon_{\phi}^{4}}(-4m_{A}^{2}m_{H}^{2}+4m_{A}^{2}s-m_{H}^{2}s)\nonumber \\
 & +\frac{\left|R\right|^{2}}{9\upsilon_{\phi}^{4}}(12m_{A}^{4}-4m_{A}^{2}s+s^{2}),\nonumber \\
Q_{4} & =\frac{4m_{A}^{2}q^{4}}{9\upsilon_{\phi}^{4}\left(2m_{H}^{2}-s\right)}(8m_{A}^{4}-8m_{A}^{2}m_{H}^{2}+8m_{A}^{2}s+2m_{H}^{4}-4m_{H}^{2}+s^{2})+\frac{4m_{A}^{2}q^{2}\Re(R)}{9\upsilon_{\phi}^{4}\left(2m_{H}^{2}-s\right)}(4m_{A}^{2}m_{H}^{2}\nonumber \\
 & -2m_{A}^{2}s+2m_{H}^{2}s-s^{2}),\,Q_{5}=-\frac{2q^{4}m_{A}^{4}}{3\upsilon_{\phi}^{4}},\,Q_{6}=\frac{4q^{4}m_{A}^{6}}{9\upsilon_{\phi}^{4}\left(2m_{H}^{2}-s\right)}.
\end{align}
with $\left\{ q,R\right\} =\left\{ s_{\beta},\frac{c_{\beta}\rho_{2}\upsilon_{\phi}}{s-m_{\eta}^{2}+im_{\eta}\Gamma_{\eta}}-\frac{s_{\beta}\rho_{h}\upsilon_{\phi}}{s-m_{h}^{2}+im_{h}\Gamma_{h}}+s_{\beta}^{2}\right\} $
for $h$; and $\left\{ q,R\right\} =\left\{ -c_{\beta},\frac{s_{\beta}\rho_{\eta}\upsilon_{\phi}}{s-m_{\eta}^{2}+im_{\eta}\Gamma_{\eta}}+\frac{c_{\beta}\rho_{1}\upsilon_{\phi}}{s-m_{h}^{2}+im_{h}\Gamma_{h}}+c_{\beta}^{2}\right\} $
for $\eta$, where $\Gamma_{h}$ ($\Gamma_{\eta}$) is the total decay
width of the Higgs (scalar $\eta$).

For the final state $h\eta$, the parameters are given by 
\begin{align}
A & =\frac{m_{h}^{2}+m_{\eta}^{2}-s}{2m_{A}^{2}},\,B=\frac{1}{2m_{A}^{2}}\sqrt{1-\frac{4m_{A}^{2}}{s}}\sqrt{(s-m_{h}^{2}-m_{\eta}^{2})^{2}-4m_{h}^{2}m_{\eta}^{2}},\nonumber \\
Q_{0} & =\frac{\sqrt{(s-m_{h}^{2}-m_{\eta}^{2})^{2}-4m_{h}^{2}m_{\eta}^{2}}}{16\pi\,s},\,R=-\frac{s_{\beta}\rho_{2}\upsilon_{\phi}}{s-m_{h}^{2}+im_{h}\Gamma_{h}}+\frac{c_{\beta}\rho_{1}\upsilon_{\phi}}{s-m_{\eta}^{2}+im_{\eta}\Gamma_{\eta}}-c_{\beta}s_{\beta},\nonumber \\
Q_{1} & =\frac{s_{2\beta}^{2}}{36m_{A}^{4}\upsilon_{\phi}^{4}}(48m_{A}^{8}-16m_{A}^{6}m_{\eta}^{2}-16m_{A}^{6}m_{h}^{2}+8m_{A}^{4}m_{\eta}^{4}+8m_{A}^{4}m_{\eta}^{2}m_{h}^{2}-8m_{A}^{4}m_{\eta}^{2}s+8m_{A}^{4}m_{h}^{4}-8m_{A}^{4}m_{h}^{2}s\nonumber \\
 & +4m_{A}^{4}s^{2}-4m_{A}^{2}m_{\eta}^{4}m_{h}^{2}-4m_{A}^{2}m_{\eta}^{2}m_{h}^{4}+4m_{A}^{2}m_{\eta}^{2}m_{h}^{2}s+m_{\eta}^{4}m_{h}^{4}),\nonumber \\
Q_{2} & =\frac{s_{2\beta}^{2}}{18m_{A}^{2}\upsilon_{\phi}^{4}}(-4m_{A}^{4}m_{\eta}^{2}-4m_{A}^{4}m_{h}^{2}+8m_{A}^{4}s+8m_{A}^{2}m_{\eta}^{2}m_{h}^{2}-2m_{A}^{2}m_{\eta}^{2}s-2m_{A}^{2}m_{h}^{2}s-m_{\eta}^{4}m_{h}^{2}-m_{\eta}^{2}m_{h}^{4})\nonumber \\
 & +\frac{\Re(R)s_{2\beta}}{9m_{A}^{2}\upsilon_{\phi}^{4}}(-24m_{A}^{6}+4m_{A}^{4}m_{\eta}^{2}+4m_{A}^{4}m_{h}^{2}+4m_{A}^{4}s-2m_{A}^{2}m_{\eta}^{4}+2m_{A}^{2}m_{\eta}^{2}m_{h}^{2}+2m_{A}^{2}m_{\eta}^{2}s-2m_{A}^{2}m_{h}^{4}\nonumber \\
 & +2m_{A}^{2}m_{h}^{2}s-2m_{A}^{2}s^{2}-m_{\eta}^{2}m_{h}^{2}s),\nonumber \\
Q_{3} & =\frac{\left|R\right|^{2}}{9\upsilon_{\phi}^{4}}(12m_{A}^{4}-4m_{A}^{2}s+s^{2})+\frac{s_{2\beta}\Re(R)}{9\upsilon_{\phi}^{4}}(2m_{A}^{2}m_{\eta}^{2}+2m_{A}^{2}m_{h}^{2}-4m_{A}^{2}s+m_{\eta}^{2}s+m_{h}^{2}s)\nonumber \\
 & +\frac{s_{2\beta}^{2}}{36\upsilon_{\phi}^{4}}(8m_{A}^{4}-4m_{A}^{2}m_{\eta}^{2}-4m_{A}^{2}m_{h}^{2}+4m_{A}^{2}s+4m_{\eta}^{2}m_{h}^{2}+m_{\eta}^{4}+m_{h}^{4}),\nonumber \\
Q_{4} & =-\frac{s_{2\beta}\Re(R)m_{A}^{2}}{9\upsilon_{\phi}^{4}}(2m_{A}^{2}+s)-\frac{s_{2\beta}^{2}m_{A}^{2}}{18\upsilon_{\phi}^{4}}(m_{\eta}^{2}+m_{h}^{2}),\,Q_{5}=\frac{s_{2\beta}^{2}}{36}\frac{m_{A}^{4}}{\upsilon_{\phi}^{4}},\,Q_{6}=0.
\end{align}

For the final states $Ah$ and $A\eta$, the parameters are given
by 
\begin{align}
A & =\frac{m_{A}^{2}+m_{H}^{2}-s}{2m_{A}^{2}},\,B=B=\frac{1}{2m_{A}^{2}}\sqrt{1-\frac{4m_{A}^{2}}{s}}\sqrt{(s-m_{H}^{2}-m_{A}^{2})^{2}-4m_{H}^{2}m_{A}^{2}},\nonumber \\
Q_{0} & =\frac{\sqrt{(s-m_{H}^{2}-m_{A}^{2})^{2}-4m_{H}^{2}m_{A}^{2}}}{16\pi\,s},\nonumber \\
Q_{1} & =-\frac{q^{2}}{36m_{A}^{4}\upsilon_{\phi}^{4}}(20m_{A}^{8}+588m_{A}^{6}m_{H}^{2}+256m_{A}^{6}s-461m_{A}^{4}m_{H}^{4}-352m_{A}^{4}m_{H}^{2}s-96m_{A}^{4}s^{2}\nonumber \\
 & +72m_{A}^{2}m_{H}^{6}+108m_{A}^{2}m_{H}^{4}s+128m_{A}^{2}m_{H}^{2}s^{2}-36m_{H}^{4}s^{2}),\nonumber \\
Q_{2} & =-\frac{q^{2}}{18m_{A}^{2}\upsilon_{\phi}^{4}(s-m_{A}^{2})}(336m_{A}^{8}-165m_{A}^{6}m_{H}^{2}-264m_{A}^{6}s-37m_{A}^{4}m_{H}^{4}+241m_{A}^{4}m_{H}^{2}s\nonumber \\
 & +128m_{A}^{4}s^{2}-21m_{A}^{2}m_{H}^{4}s-88m_{A}^{2}m_{H}^{2}s^{2}-32m_{A}^{2}s^{3}-18m_{H}^{4}s^{2}+36m_{H}^{2}s^{3}),\nonumber \\
Q_{3} & =\frac{q^{2}}{36\upsilon_{\phi}^{4}(s-m_{A}^{2})^{2}}(-591m_{A}^{8}+480m_{A}^{6}m_{H}^{2}-302m_{A}^{6}s+m_{A}^{4}m_{H}^{4}-416m_{A}^{4}m_{H}^{2}s\nonumber \\
 & +189m_{A}^{4}s^{2}+6m_{A}^{2}m_{H}^{4}s+104m_{A}^{2}m_{H}^{2}s^{2}-100m_{A}^{2}s^{3}+9m_{H}^{4}s^{2}-72m_{H}^{2}s^{3}+36s^{4}),\nonumber \\
Q_{4} & =\frac{q^{2}m_{A}^{2}}{18\upsilon_{\phi}^{4}(s-m_{A}^{2})^{2}}(-153m_{A}^{6}-69m_{A}^{4}m_{H}^{2}+164m_{A}^{4}s-2m_{A}^{2}m_{H}^{2}s-29m_{A}^{2}s^{2}-9m_{H}^{2}s^{2}+18s^{3}),\nonumber \\
Q_{5} & =\frac{q^{2}m_{A}^{4}}{36\upsilon_{\phi}^{4}(s-m_{A}^{2})^{2}}(153m_{A}^{4}-2m_{A}^{2}s+9s^{2}),\,Q_{6}=0,
\end{align}
with $\{m_{H},q^{2}\}=\{m_{h},s_{\beta}\},\{m_{\eta},-c_{\beta}\}$
for the final states $Ah$ and $A\eta$, respectively.

\end{document}